\documentclass{ametsocV6.1nolinenumbers}

\title{Abyssal Slope Currents}

\authors{Esther Cap{\'o}\aff{a},\correspondingauthor{Esther Cap{\'o},
    estherct@ucla.edu} James C. McWilliams\aff{a},
  Jonathan Gula\aff{a,b}, M. Jeroen Molemaker\aff{a}, Pierre
  Damien\aff{a}, Ren{\'e} Schubert\aff{b} }
\affiliation{\aff{a}{Department of Atmospheric and Oceanic Sciences,
    University of California, Los Angeles, CA, USA}\\
\aff{b}{Univ Brest, CNRS, IRD, Ifremer, Laboratoire d’Oc\'eanographie 
Physique et Spatiale (LOPS), IUEM, Brest, France}
}

\abstract{Realistic computational simulations in different oceanic 
basins reveal prevalent prograde mean flows
  (in the direction of topographic Rossby wave propagation along
  isobaths; a.k.a. topostrophy) on topographic slopes in the deep
  ocean, consistent with the barotropic theory of eddy-driven mean
  flows.  Attention is focused on the Western
  Mediterranean Sea with strong currents and steep topography. 
  These prograde mean currents induce an opposing
  bottom drag stress and thus a turbulent boundary-layer mean flow in
  the downhill direction, evidenced by a near-bottom negative mean
  vertical velocity. The slope-normal profile of diapycnal buoyancy mixing results in 
  down-slope mean advection near the bottom (a tendency to locally 
  increase the mean buoyancy) and up-slope buoyancy mixing (a
  tendency to decrease buoyancy) with associated buoyancy fluxes
  across the mean isopycnal surfaces (diapycnal
  downwelling).  In the upper part of the boundary layer and nearby
  interior, the diapycnal turbulent buoyancy flux divergence reverses
  sign (diapycnal upwelling), with upward Eulerian mean buoyancy
  advection across isopycnal surfaces.  These near-slope
  tendencies abate with further distance from the boundary.  An
  along-isobath mean momentum balance shows an advective acceleration
  and a bottom-drag retardation of the prograde flow.  The eddy
  buoyancy advection is significant near the slope, and the associated
  eddy potential energy conversion is negative, consistent with mean
  vertical shear flow generation for the eddies.  This cross-isobath
  flow structure differs from previous proposals, and a new
  one-dimensional model is constructed for a topostrophic, stratified,
  slope bottom boundary layer.  The broader issue of the return
  pathways of the global thermohaline circulation remains open, but
  the abyssal slope region is likely to play a dominant role.}

\begin{document}
\def\nab{\mathbf{\nabla}}
\def\pd{\partial}
\def\ie{{\it i.e., }}
\def\eg{{\it e.g., }}
\def\etc{{\it etc.}}
\def\etal{{\it et al.}}
\def\viz{{\it viz., }}
\def\vs{{\it vs. }}
\def\nb{{\it n.b., }}
\def\cf{{\it cf., }}
\def\la{\langle\,}
\def\ra{\,\rangle}
\def\bu{{\bf u}}
\def\bx{{\bf x}}
\def\cblue#1{{\color{blue}#1}} 

\maketitle

\section{Introduction}
\label{sec:intro}
The abyssal ocean below the pycnocline is little explored relative to
the upper ocean.  It is understood to be generally full of mesoscale
eddy currents that reach deeply down from the pycnocline,
inertia-gravity waves generated at both the bottom and surface,
stratified shear turbulence, a turbulent bottom boundary layer with
topographic vortical wakes extending into the interior, submesoscale
coherent vortices, and narrow deep western boundary currents that are
part of the global thermohaline circulation.  This paper is about
additional phenomena that as yet are even less well understood.

The first of these phenomena is topostrophy \citep{holl2008b}, defined
as a prograde time-mean along-isobath current over a sloping bottom 
(\ie in the same direction as the phase propagation of topographic
Rossby waves, with shallower water to the right of the current in the
Northern Hemisphere where the Coriolis frequency $f$ is positive, and
\textit{vice versa} in the Southern Hemisphere).  Its measure is defined
by
\begin{equation}
  \mathcal{T}\,=\, \mathrm{sgn}[f] \ (\widehat{\vec{z}} \cdot \bu
  \times \nabla h) \,,
\label{topos}
\end{equation}
and its time average near the bottom, denoted by
$\overline{\mathcal{T}}_b$, is usually positive.  The subscript $b$
denotes a near-bottom value (\ie within the Monin-Obukhov turbulent
layer, outside of any viscous sublayer that would support a no-slip
boundary condition).  The overbar is a time average; bold-face symbols
are horizontal vectors; overlying arrows indicate a 3D vector;
$\widehat{\vec{z}}$ is a unit vertical vector in the upward direction
against gravity; and $h(\bx)$ is the resting depth of the ocean.  A
unit horizontal vector in the topostrophic direction is
\begin{equation}
  \widehat{\mathbf{s}} \,=\, \mathrm{sgn}[f] \ \frac{\nabla h}{|\nabla h|}\,
  \times\, \hat{\vec{z}} \,,
\label{sdef}
\end{equation}
whence $\mathcal{T} \,=\, \widehat{\mathbf{s}} \cdot \bu \ |\nabla h|$.

The theory for this phenomenon has its origins in topographic
turbulence theory for two-dimensional flow over variable bathymetry
\citep{brha1976,salm1976,herr1977}.  It shows that random eddying
flows evolve to develop a negative correlation between the vertical
vorticity, $\zeta = \pd_xv-\pd_yu$ where $(u,v)$ are
(eastward,northward) horizontal velocities, and resting-depth
variations, $h - H$, where $H$ is a mean depth.  This implies the
development of a persistent topostrophic flow that can be described as
an eddy-driven mean flow.  The rationales for this outcome have
variously been given as viscous enstrophy dissipation due to the
forward cascade of $\overline{\zeta^{\prime\,2}}$ in two-dimensional
turbulence, entropy maximization in the sense of statistical
mechanical equipartition of evolutionary end-states, and potential
vorticity homogenization through horizontal eddy stirring.
Topostrophy is also called the Neptune effect, and Greg Holloway has
been a strong advocate for its importance
\citep[\eg][]{holl1987,holl1996}.  For oceanic basins, the most
energetic eddies are mesoscale, and the largest $|\nabla h|$ values
occur over continental slopes, seamounts, mid-ocean ridges, and
islands, and this is where topostrophy might be most likely to occur;
such flows would be in the cyclonic direction on basin edges and
anticyclonic around interior bumps.  Some striking examples have been
reported \citep{alvarez1994effect,saki1995,weij2020,xie2022}, and a climatological data
set from bottom current meters supports the frequent occurrence of
topostrophy on slopes \citep{holl2008b}.  In a companion paper
\citep{schu2024} and in this one, high-resolution, realistic
simulations are used to demonstrate the extreme prevalence of
topostrophic flows on abyssal slopes.

The second phenomenon focused on in this paper is a prevalent downhill
mean flow.  If the along-isobath flow is topostrophic, then there is a
horizontal drag stress on the bottom, $\mathbf{\tau} \,=\, \rho C_d
|\overline{\bu}_b|\overline{\bu}_b$, in the same direction, acting to
retard the near-bottom $\overline{\bu}_b$.  This stress could cause a
turbulent bottom (Ekman) boundary layer, whose mean horizontal
transport velocity --- ignoring the influence of stable density
stratification and buoyancy mixing (but see Sec.~\ref{sec:1D}) --- is
perpendicular to the bottom stress in the downhill direction, \ie
$\overline{\bu}_b \cdot \nabla h \,>\, 0$.  Alternatively expressed,
the unstratified Ekman layer transport is to the left of the adjacent
interior flow in the Northern Hemisphere (and \textit{vice versa} in
the South).  Furthermore, the mean kinematic bottom boundary condition
of no normal flow at a solid boundary is
\begin{equation}
\overline{w}_b \,=\, -\ \overline{\bu}_b \cdot \nabla h \,,
\label{wb}
\end{equation}
and this implies that the vertical velocity $\overline{w}_b \,<\,0$ at
the bottom.

\begin{figure}[t]
  \centering\includegraphics[width=19pc,angle=0]{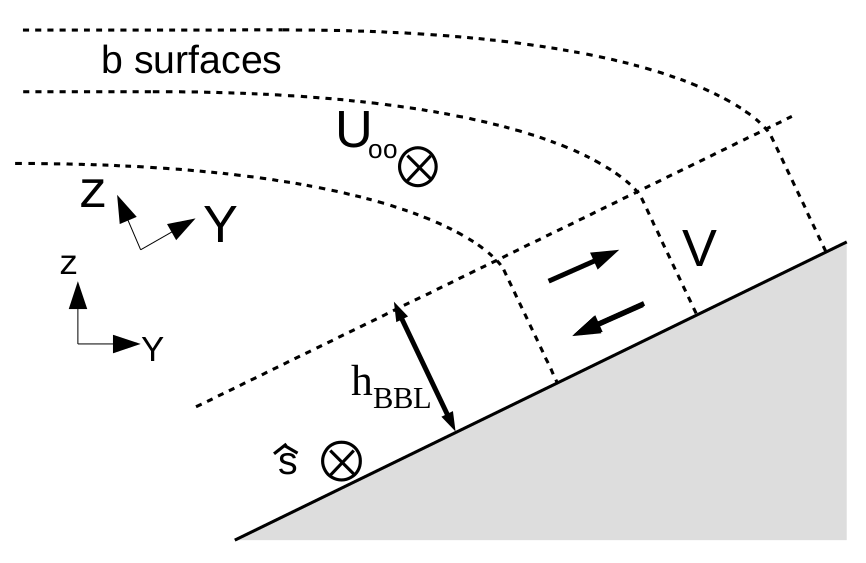}\\
\caption{Sketch of the currents and buoyancy surfaces near a sloping
  boundary for a topostrophic interior flow when $f > 0$. Y and $Z$
  are the coordinates parallel and perpendicular to the slope,
  respectively. $h_{BBL}$ is the BBL thickness, and $V(Z)$ is the
  cross-slope velocity profile within the BBL; see (\ref{XYZ}).}
\label{fig:sketch}
\end{figure}

Care must be taken about details of the vertical profile of the
current structure within the boundary layer.  In this regard, the
third primary phenomenon that we focus on is the mean diapycnal
buoyancy mixing associated with these slope flows.  An important
idealized, one-dimensional (1D) bottom boundary layer (BBL) model
\citep{wuns1970,phil1970,garr1990} for the effect of mixing in a
stratified fluid with flat buoyancy surfaces in the interior next to a
topographic slope has steady solutions with an upslope mean current at
the bottom (hence $\overline{w}_b \,>\, 0$), a reverse downslope
current above, and a retrograde along-isobath mean current in the
adjacent interior --- all as a consequence of no buoyancy flux through
the boundary (ignoring the small geothermal heat flux \citep{thom1996}); see
Fig.~\ref{fig:sketch}.  Stable stratification is expressed as $\pd_z b
\,>\, 0$, where $b \,=\, - g\rho/\rho_0$ is the buoyancy, $\rho$ is
the local potential density, and $z$ is the upward coordinate aligned
with gravity.  The direction of these currents is in contradiction to
the topostrophic and Ekman-layer expectations in the preceding two
paragraphs.  This contradiction is resolved in Sec.~\ref{sec:1D}.

The small-scale turbulent buoyancy flux,
\begin{equation}
  \overline{\vec{\mathcal{B}}} \,=\, \overline{\vec{u}'b'} \,,
\end{equation}
(where primes denote deviations from mean values that must be
  defined in each context; in this case, deviations include sub-grid
  scale fluctuations due to parameterizations and model discretization
  errors)
is ``downward'' in stable stratification, \ie toward less buoyant
water (\eg by mixing warm surface water and abyssal cold water).
Extensive micro-structure measurements
\citep{stla2001,nave2004,wate2014,gara2019} show that in the abyss
above ``rough'' topography this mixing flux is often largest at a
depth somewhat above the sloping bottom due to both boundary-layer
turbulent mixing and nearby breaking of bottom-generated
inertia-gravity waves or topographic wakes.  The diapycnal mixing
velocity across a buoyancy iso-surface is defined by
\begin{equation}
  \widetilde{\omega} \,=\, - \
  \frac{\vec{\nabla} \cdot \overline{\vec{\mathcal{B}}}}
       {|\vec{\nabla}\overline{b}|} \,.
\label{omega}
\end{equation}
Integrated over the horizontal area of a buoyancy surface in the
abyss, $\widetilde{\omega}$ must be positive both to serve as the upward return flow compensating polar sinking in the global thermohaline circulation, and to account for the water mass transformation of less buoyant into more buoyant water.  Invoking the slope buoyancy-mixing BBL model cited above and
the observed vertical structure of diapycnal mixing near topography
\citep[\eg][]{ferr2016,mcfe2017,call2018,homc2020,drak2022,bake2023},
it is hypothesized that most of the dominant positive
$\widetilde{\omega}$ occurs in the BBL next to the slopes and weaker
negative values occur in the interior, with a spatially inhomogeneous
pattern concentrated near topographic slopes.  This is a modern
revision of the older conceptions of positive $w$ and
$\widetilde{\omega}$ more homogeneously spread throughout the abyssal
interior \citep{stom1958,munk1966}.

The intersection of these processes occurs principally in
the mean buoyancy balance,
\begin{align}
\pd_t \overline{b} &= \ - \ \overline{\vec{u} \cdot \vec{\nabla}b} 
\ -\ \vec{\nabla} \cdot \overline{\vec{\mathcal{B}}} \\
&= -\ \overline{\vec{u} \cdot \vec{\nabla}b} \ + \
\widetilde{\omega} \, |\vec{\nabla}\overline{b} |
\,\approx\, 0 \,,
\label{mean-bal}
\end{align}
especially near bottom slopes.  The mean and eddy buoyancy advection
approximately balances the small-scale mixing, \ie their tendencies
must have opposite signs in equilibrium.

In this paper we partially synthesize these processes guided by
realistic regional simulations, but leaving the full characterization
of the upward branches of the thermohaline circulation to future
global simulations.  Our findings are that topostrophic flows are very
common and that $\overline{w}$ and $\widetilde{\omega}$ are mostly
negative along the abyssal slopes, but with sign reversals toward the
interior.  Our focus is on the abyss because there can be many
competing influences on the mean currents in the upper ocean (\eg the
wind-driven western boundary current in a subtropical gyre); whereas,
mesoscale eddies are ubiquitous and are often the dominant influence
on the mean currents in the abyss \citep[\eg][]{rhho1979}.  We use the
Western Mediterranean Sea to illustrate this behavior.
\cite{schu2024} is a companion paper showing that this is common
throughout the Atlantic and Pacific Oceans, and \cite{gula2024}
explains the method of diagnosing $\vec{\mathcal{B}}$ in a
terrain-following coordinate model (as used here), inclusive of both
explicitly parameterized mixing and the contribution from model
discretization errors.

The organization of the paper is as follows.  The general circulation
of the Western Mediterranean Sea is summarized in
Sec.~\ref{sec:MedGC}; the simulation specifications are in
Sec.~\ref{sec:model}; the near-slope circulation results are in
Sec.~\ref{sec:topo-w}; the mean along-isobath momentum 
$\widehat{\mathbf{s}} \cdot \overline{\bu}$ and buoyancy $\overline{b}$ 
balances are diagnosed in Secs.~\ref{sec:Mbal}-\ref{sec:Bbal}; a
simple 1D BBL model combining diapycnal mixing and a topostrophic
interior flow is in Sec.~\ref{sec:1D}; and the results are further
discussed and summarized in Secs.~\ref{sec:discuss}-\ref{sec:sum}.

\section{Western Mediterranean circulation}
\label{sec:MedGC}
Topostrophy is a widespread phenomenon \citep{holl2008b,schu2024}.
The analysis here is conducted in the Western Mediterranean basin,
with a particular focus in the Alboran Sea, partly taking advantage of
existing simulations and previous analyses
\citep{capo2021,camc2022a,capo2023}.

The Western Mediterranean Sea is a semi-enclosed basin located
between the Strait of Gibraltar, 36$^{\circ}$N--5.5$^{\circ}$W, and
the Strait of Sicily, 37$^{\circ}$N--12$^{\circ}$E
(Fig.\ref{fig:WM_To_wb}a). The region is characterized by a complex
bathymetry shaped by numerous capes, seamounts, ridges and islands of
different shapes and sizes.  While the interior of the basin is
relatively flat, with an averaged depth not exceeding 3000 m, the
continental and island/ridge margins are characterized by steep
slopes, often shaped with narrow, deep canyons. The Alboran Sea is the
transitional area between 5.5$^{\circ}$W--1$^{\circ}$W that connects
the Western Mediterranean Sea with the Atlantic Ocean through the
narrow Strait of Gibraltar (Fig.\ref{fig:WM_To_wb}a,b).

In the upper 150-200 m the large scale circulation in the Alboran Sea
is mainly driven by the incoming Atlantic Jet, a strong fresh, surface
current with velocities of $\mathcal{O}$(1) m s$^{-1}$ forced by a
strong surface pressure gradient between both sides of the Strait and
modulated by the tides affecting the Atlantic side.  This jet flows
along the basin shaping the quasi-permanent Western Anticyclonic Gyre,
and an intermittent, less intense Eastern Anticyclonic Gyre
\citep{viud1996,varg2002}. To the east, the meeting of the Atlantic
fresh current with the resident salty Mediterranean water generates a
strong surface density (mainly salinity) contrast, commonly named the
Almeria-Oran Front \citep{tint1988,alle2001}.  The Atlantic Jet exits
the Alboran Sea and flows eastwards along the African coast towards
the Strait of Sicily as the Algerian Current.  The Mediterranean
thermohaline circulation transforms the incoming fresh Atlantic water
into saltier water, which leaves the basin towards the Atlantic Ocean
through the Strait of Gibraltar in the form of Levantine Intermediate
Waters, within a 200-600 m depth range on average, and Deep
Mediterranean Water below.  These currents follow a complicated path
that is highly constrained by the shape of topography: a major branch
of the Levantine undercurrent flows along the Spanish margin towards
the Strait, while the denser Deep Mediterranean Water occupies the
central (and deepest) part of the basin and finally flows towards the
Strait along the African slope \citep{sanc2022}. The influence of 
sub-basin scale interactions between oceanic eddies and 
bottom topography in the western Mediterranean Sea was first investigated in 
\citet{alvarez1994effect}. By testing a parameterization of the eddy-topography interaction through various numerical experiments, their study suggested that topographic 
stress could play a significant role in driving circulation at both basin and sub-basin scales.

The Western Mediterranean Sea exhibits a high degree of eddy
variability, manifested in its boundary currents, surface density
fronts, pycnocline depth variations, large and rapid vertical parcel
displacements, topographic vorticity generation, and abyssal currents,
\eg partly as shown in the recent CALYPSO experiments
\citep{maha2020}. 

\section{Mediterranean simulations}
\label{sec:model}
We use two realistic simulations with the Regional Oceanic Modeling
System (ROMS) \citep{shmc2005} and different spatial grid resolutions:
(1) the Western Mediterranean basin (WMED) with horizontal $dx = 1500$
m and 60 $\sigma$-levels in the vertical (with stretching parameters
$\theta_{s}$ = 7, $\theta_{b}$ = 5 near the surface and the bottom
respectively, following \citet{lema2012a}); and (2) the nested
subdomain for the Alboran Sea (ALB) with $dx = 500$ m, 140
$\sigma$-levels, and $\theta_{s} = \theta_{b} = 7$.  WMED was
calculated using CROCO, a modeling system built upon ROMS with a
numerical kernel close to the UCLA version of the ROMS model
\citep[][\textit{http://www.croco-ocean.org}]{shmc2009a}.
ALB was calculated using the UCLA-ROMS version of the model
(\textit{https://github.org/CESR-lab/ucla-roms.git}).  The two codes
have enough of a common heritage that their solution behaviors are
similar.

ROMS is a split-explicit, 3D free-surface oceanic model that solves
the hydrostatic Primitive Equations for momentum, temperature, and
salinity using the Boussinesq approximation and hydrostatic vertical
momentum balance \citep{shmc2005}.  Horizontal advection is based on a
modified version of the upstream 3rd-order scheme, which is split into
a 4th-order centered advection and a rotated biharmonic diffusion with
grid-dependent diffusivity satisfying the Peclet constraint
\citep{marc2009,lema2012a}.  To avoid numerical instabilities, the
algorithm for vertical advection of momentum combines a high-order
explicit scheme with an implicit one, which mitigates the vertical
Courant-Friedrichs-Lewy restriction \citep{shch2015}.
Vertical advection of tracers is performed using a a 4th-order Akima
scheme.

In both configurations the top and bottom boundary layers are well
resolved using a K-Profile Parameterization (KPP) for the turbulent
mixing \citep{larg1994}.  With this scheme the mixing coefficients
reach maximum values in the middle of the layers, decreasing to
minimal background values within the interior, except where the
Richardson number, $Ri$, becomes small.  Bottom drag is parameterized
using the quadratic law, $\tau = \rho_{0} C_{D} \Vert \textbf{u}_{b}
\Vert \textbf{u}_{b}$, where $\rho_{0}$ is the reference density,
$\textbf{u}_{b}$ is the horizontal velocity in the bottom layer, and
$C_{D}$ is a drag coefficient.  We use the logarithmic formulation for
$C_{D} = [k / log \left( \Delta z_{b} / z_{r} \right)]^{2}$, where $k
= 0.41$ is the von Karman constant, $\Delta z_{b}$ is the thickness of
the bottom layer, and $z_{r}$ is the roughness scale, set to $1$ cm.

Bathymetry for both domains is obtained from the SRTM30$\_$PLUS$\_$V11
global dataset \citep{beck2009}.  A steepness factor $r$ $<$
0.2 is applied to realistic topography to limit pressure
gradient errors. An additional smoothing with a radius of 4 km
half-width is also applied to the grid in the ALB solution.

Realistic atmospheric forcing for the WMED solution is derived
  from the NCEP-CFSR data set at 0.2$^\circ$ resolution. Wind
  velocity, temperature and humidity fluxes, and short and long wave
  radiation fluxes feed the model at hourly time steps, and river
  inflow is derived from a climatological dataset.  The initial and
  boundary conditions for WMED are interpolated from the Copernicus
  Marine Service global ocean forecasting system \citep{lell2018},
  providing a realistic representation of the Western Mediterranean
  Sea in the mesoscale range \citep{maso2019}.  This simulation was
  run for four years, 2010-2011, with a spin-up period of
  approximately three months, although the first year is excluded from
  the analysis. \citet{capo2020} provides a detailed description
  of this solution, including validation and data analysis.

The corresponding atmospheric forcing for ALB is obtained from
  the ECMWF atmospheric reanalysis of the global climate product
  (ERA5) at hourly time steps \citep{era5}.  The initial state and
  boundary conditions for ALB are obtained from WMED, and the
  simulation is initialized in December 2010, eleven months after WMED
  initialization to ensure sufficient spin-up of the latter.

Boundary conditions consist of a Flather-type scheme for the
barotropic mode, and Orlansky-type for baroclinic velocity and for
tracers, following the nesting procedure described in
\citet{maso2010}.

Neither of the simulations analized here is forced with
  tides. Although internal tides play a crucial role in enhancing
  mixing over abyssal slopes in the oceans, these are not the sole
  contributors in our study region and are not so essential for our
  analysis. Lee waves, topographic wakes, submesoscale instabilities
  and interactions with mesoscale eddies also play important roles in
  enhancing mixing along abyssal slopes \citep{legg2021mixing, mashayek2024role}, and these processes occur in
  our solution at submesoscale resolution despite the absence of tidal
  forcing.  While the tidal effect is important as part of the
  Atlantic inflow through the Strait of Gibraltar, tidal amplitudes
  decay towards the interior of the basin, and the effect of
  topographic wakes, lee waves and other instability sources can be as
  relevant as internal tides in enhancing mixing near abyssal slopes
  in the Mediterranean Sea.  Comparisons with tidal solutions show
  modest differences in the topostrophy and downhill mean flows that
  are the primary focus in this paper (Sec.~\ref{sec:topo-w}).

The ALB simulation is based on a previous solution with
  identical configuration that included tidal forcing, ALB$\_$T, and
  is extensively validated and analyzed in \citet{capo2020}.
  The new ALB solution presented here shows consistency with the
  parent WMED and with ALB$\_$T. ALB validation entailed comparison of
  the 3D flow structures, temperature and salinity fields, and the
  analysis of time series of area averaged kinetic energy and local
  sea surface height throughout the domain. The latter are consistent
  with those obtained from the parent WMED and serve to determine the
  spin up period required until the model stabilizes. The absence of
  tidal forcing facilitates rapid stabilization, which is achieved
  after two weeks. For greater reliability, the first two months of
  the ALB simulation are excluded from the analyzed period.

The results presented in this paper, if not otherwise
  indicated, are obtained from daily-averaged fields of the last 3
  years of the WMED solution (2011--2013) and from the last year of
  ALB (Feb 2011-Jan 2012).

\section{Abyssal topostrophy and downhill flow}
\label{sec:topo-w}
\begin{figure}[t]
{\includegraphics[width=\textwidth,angle=0]{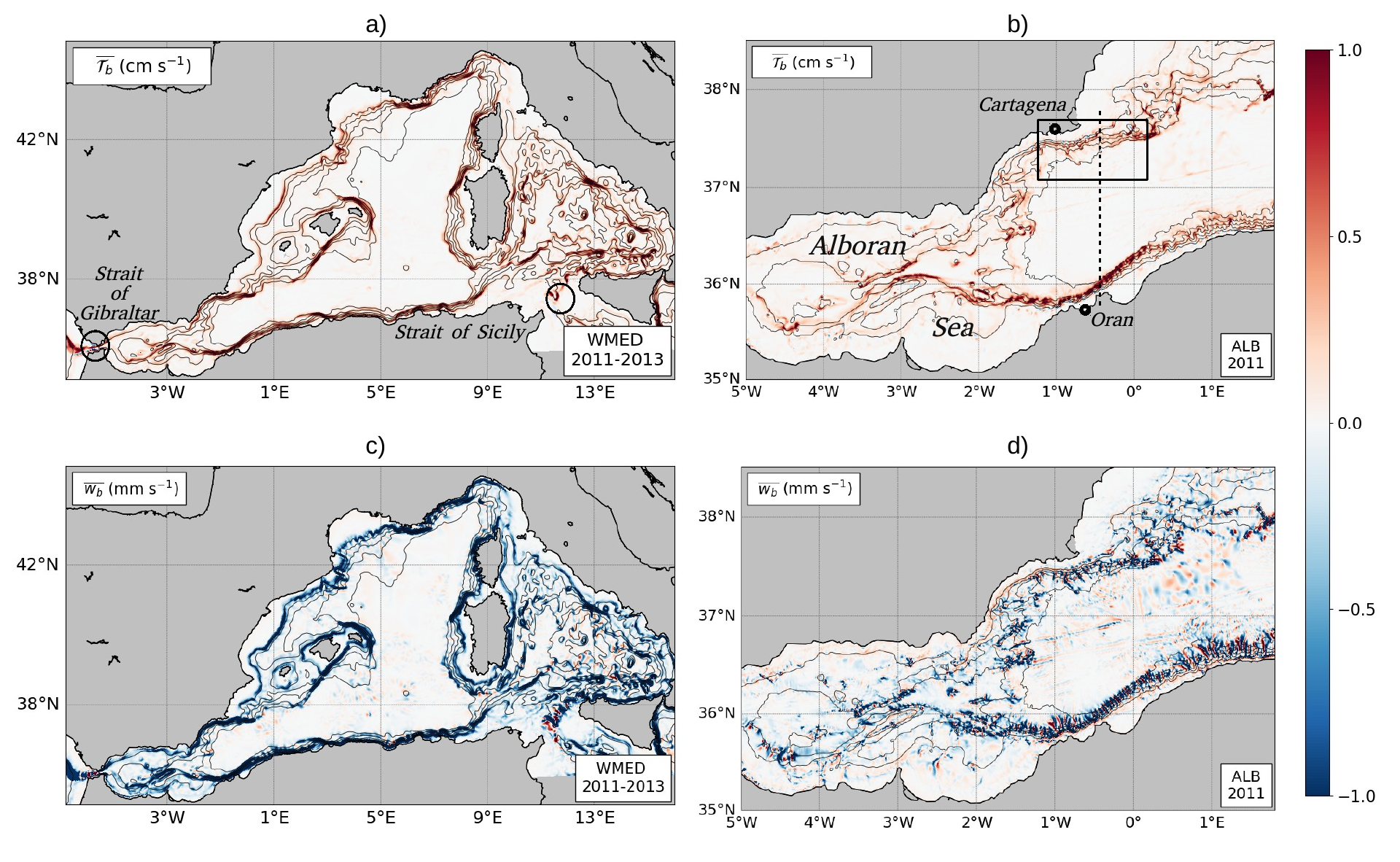}}
\caption{Time-averaged $\overline{\mathcal{T}}_b$ (a,b) and
  $\overline{w}_b$ (c,d) in the Western Mediterranean (a,c) and
  Alboran (b,d) simulations.  Bathymetry is contoured in black every
  500 m.  The black box in (b) frames the Cartagena subdomain analyzed in
  Fig.~\ref{fig:Cartagena} \textit{et seq.}, and the dashed line indicates
  the section shown in Fig.~\ref{fig:K_section}. Locations of interest 
  are labeled in (a) and (b).}
\label{fig:WM_To_wb}
\end{figure}

In this section we examine the first two phenomena exposed in Sec.~1,
topostrophy, $\overline{\mathcal{T}}$, and downhill bottom flow,
$\overline{w}_b$. using our two simulations in the Western
Mediterranean, WMED and ALB, and the definitions in
(\ref{topos})-(\ref{wb}).  Figure~\ref{fig:WM_To_wb} shows these fields
over the whole domains of WMED and ALB.  As anticipated,
$\overline{\mathcal{T}_b}$ is primarily positive, and $\overline{w}_b$
is primarily negative.  Recall that both quantities are weighted by
$\nabla h$ and thus tend to be largest over slopes.  Not all slope regions
have these signatures, but there is very little evidence of opposite
signs with significant magnitude.

\begin{figure}[ht!]  
\centering
    {\includegraphics[scale=0.6]{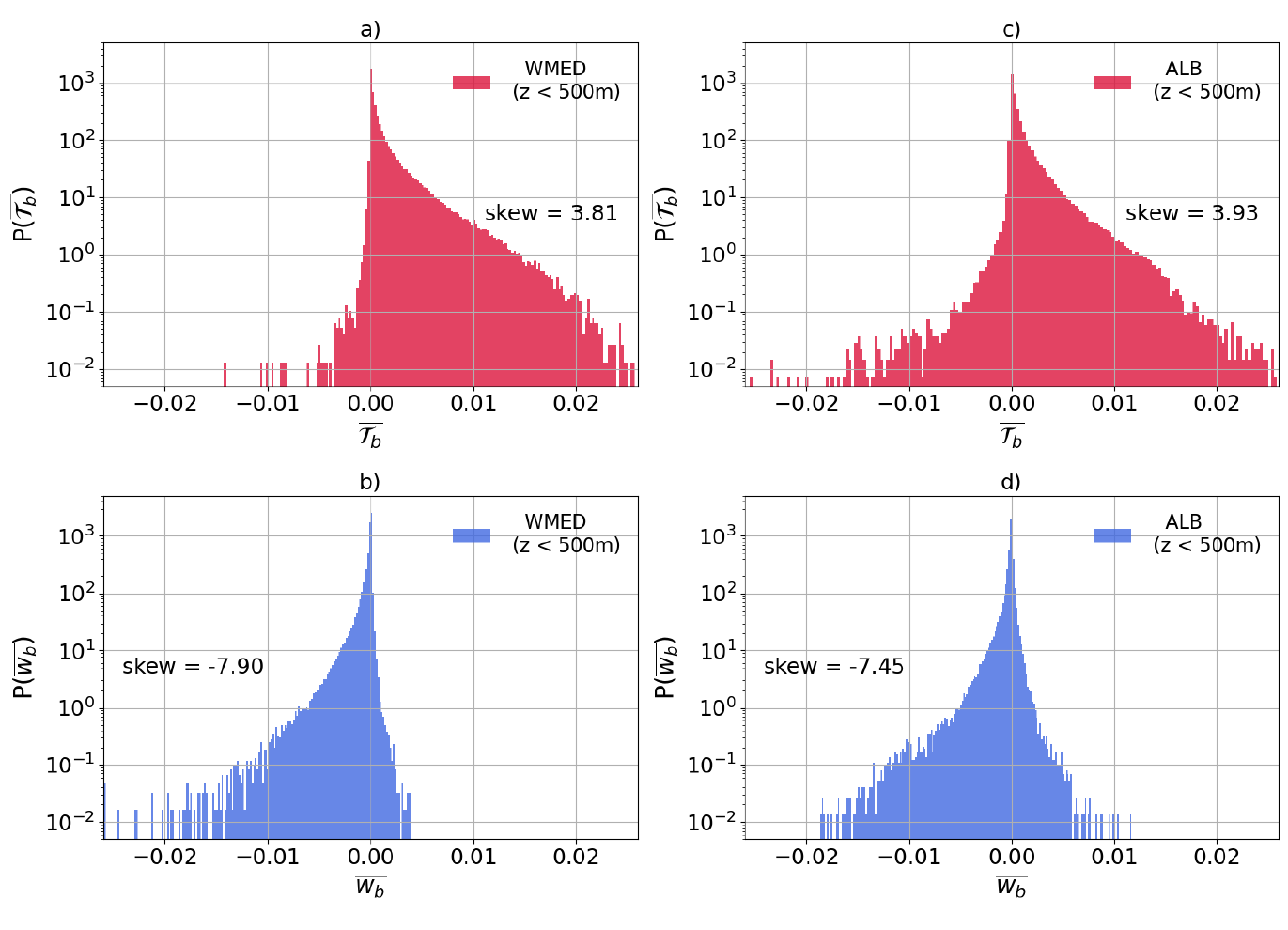}}
    \caption{Probability Density Functions for the bottom fields,
      $\overline{\mathcal{T}}_b$ [m/s] (a,c) and $\overline{w}_b$
      [m/s] (b,d), for time-averaged values at each location in the
      Western Mediterranean (a,b) and Alboran (c,d) simulations (in m
      s$^{-1}$).  Only locations with resting depths h $\ge$ 500 m are
      included.  The predominance of negative $\overline{w}_b$ and
      positive $\overline{\mathcal{T}}_b$ is evident, \eg in the
      indicated values of skewness.}
\label{fig:PDF}
\end{figure}

A more complete view is provided by the Probability Distribution
Functions (PDFs) for these fields at locations with resting depths
$\ge$ 500 m (Fig.~\ref{fig:PDF}).  There are many small values due to
locations with small $|\nabla h|$, but the PDFs are highly skewed in
both their bulk and tails.

\begin{figure}[ht!]  
\centering
{\includegraphics[scale=0.8]{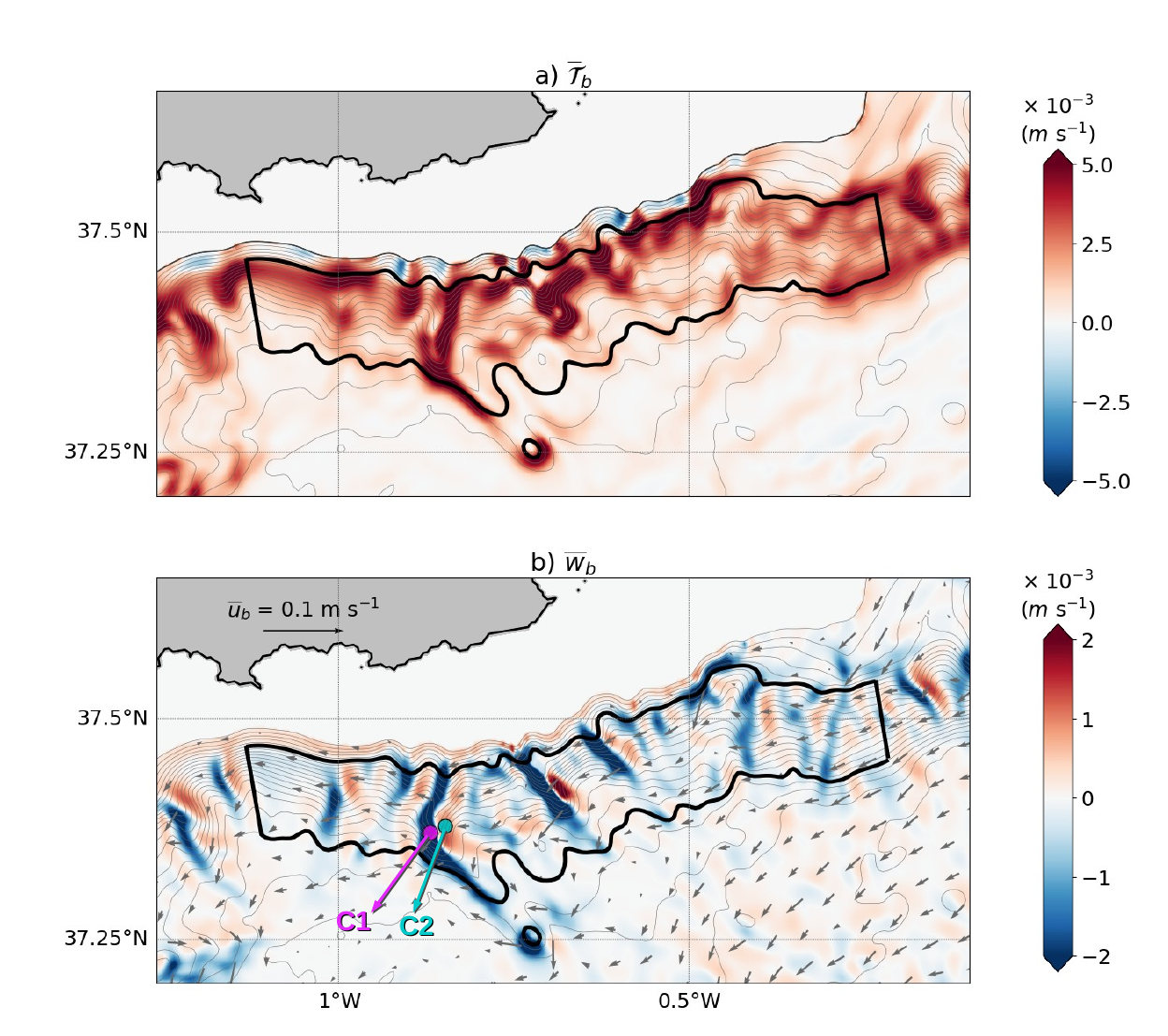}}
\caption{A zoom into the Cartagena region depicted within the black
  box in Fig.~\ref{fig:WM_To_wb}b for time-averaged
  $\overline{\mathcal{T}}_b$ (a), $\overline{w}_b$ (b; colors), and
  $\overline{\bu}_b$ (b; vectors) for resting depths $\ge$ 500 m in
  the ALB simulation.  Bathymetry is contoured in black every 100 m.
  The thick black line defines the sub-region for the along-isobath
  averages in Fig.~\ref{fig:CC_topow} \textit{et seq.}.  The points C1
  and C2 in (b) mark two representative locations on opposite sides
  along a ridge, used in Fig.~\ref{fig:canyon_ADV_KE} to illustrate
  the local flow vertical profiles.}
\label{fig:Cartagena}
\end{figure}

The ALB simulation has a significantly increased spatial bathymetry
and grid resolution compared to WMED (Sec.~\ref{sec:model}).  It
therefore exhibits greater variability in these measures, especially
in $\overline{w}_b$.  Partly, this is expressed in dipole patterns over
bumps and ridges and along the sides of submarine canyons that
interrupt the primary continental slopes.  In particular, abundant
submarine canyons interrupt the Spanish slope between
1.2$^{\circ}$W - 0.2$^{\circ}$E (the Cartagena subregion delimited in
black in Fig.~\ref{fig:WM_To_wb}) and the African margin between
2$^{\circ}$W - 1.5$^{\circ}$E (the path of the eastward Algerian
Current).  It has been widely shown from both numerical simulations
and observations that currents flowing across canyons induce
opposite-signed $\overline{w}_b$ values on the upstream (negative $w$)
and downstream (positive $w$) sides \citep{klin1996,flex2006,spal2014} 
with reversed $w$ signs for flow over bumps.  In combination with
prograde slope flow, this local canyon effect results in a net
downhill transport.  These features are shown in detail in
Fig.~\ref{fig:Cartagena} for the Cartagena subregion.  The prevailing
topostrophic, downhill flow in this regions is part of the Levantine
and Deep Mediterranean undercurrent that crosses the Alboran Sea
westward towards the Strait of Gibraltar.
We have chosen to focus on the Cartagena region, both here and below,
as one that seems at least not atypical from the perspective of
Fig.~\ref{fig:WM_To_wb}.  No doubt there are meaningful regional
differences, but based on our present level of exploration, the
behavior off Cartagena is representative of the structure and dynamics
of abyssal slope flows.

To better discern the behavior associated with the primary abyssal
slope, we average over the local flow structures evident in
Fig.~\ref{fig:Cartagena} by averaging in the $\widehat{\mathbf{s}}$
direction (\ie along isobaths) within the thick black sub-region
there.  This is done by binning all grid points along the slope with a
discretization interval of 40 m in the local depth.

\begin{figure}[ht!]  
\centering
    {\includegraphics[scale=0.45]{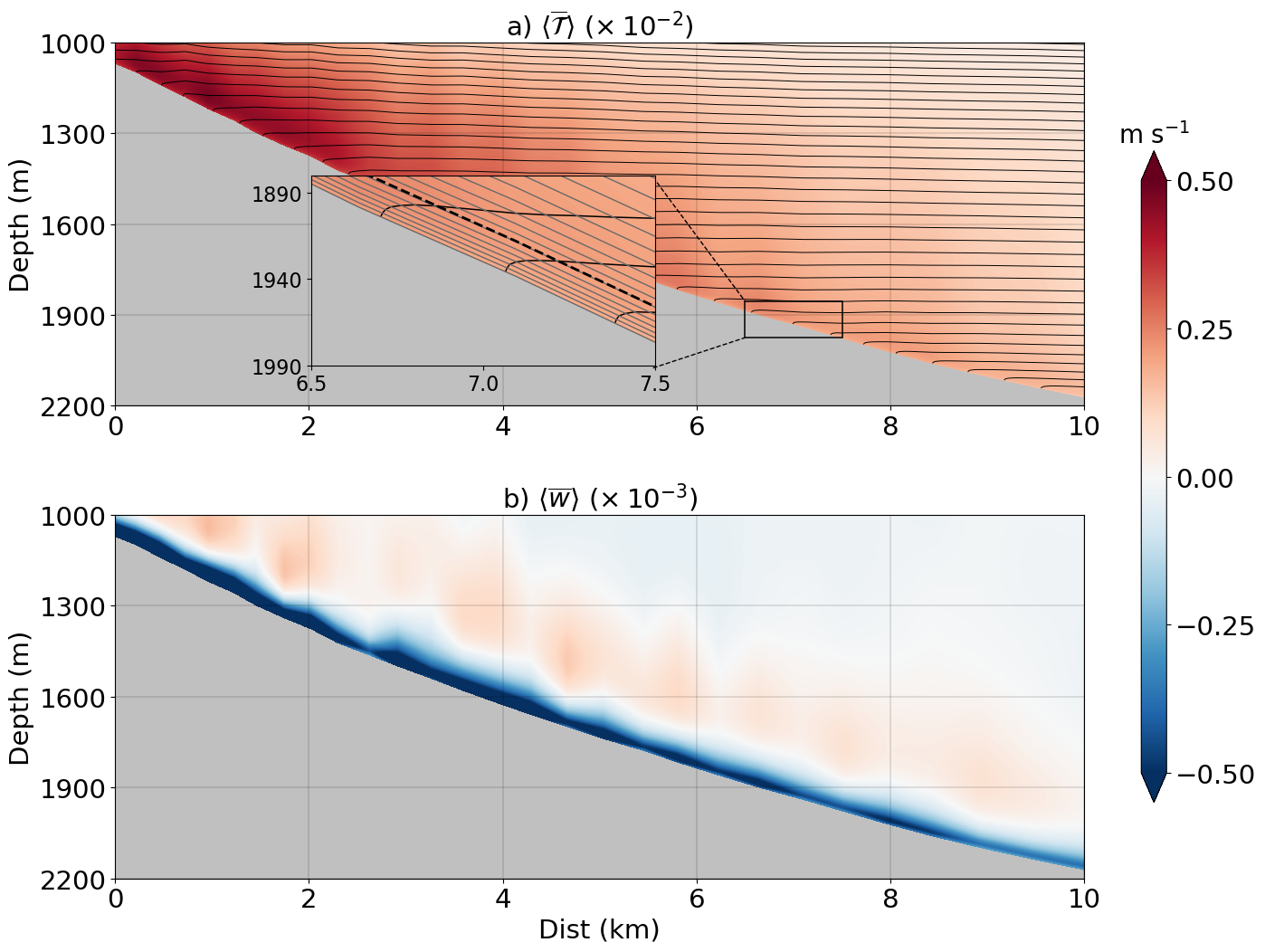}}
\caption{Time- and along-isobath-averaged cross-sections on the
  Cartagena slope (\ie within the thick black line in
  Fig.~\ref{fig:Cartagena}, between 0.25-1.1${^\circ}$W) in the ALB
  simulation: (a) topostrophy $\la \overline{\mathcal{T}}\ra(\eta,z)$
  ($\times 10^{-2}$) and (b) vertical velocity $\la \overline{w}\ra
  (\eta,z)$ ($\times 10^{-3}$), both referenced to the color bar
  values on the right side.  $\eta$ is a horizontal coordinate in the
  cross-isobath direction, increasing toward deeper water. Isopycnals
  are contoured in black in (a), and the inset box shows in detail the
  tilting of isopycnals perpendicular to the sloping floor, consistent
  with zero buoyancy flux at the boundary. Also in the inset, the
  dashed black line represents the mean bottom boundary-layer depth
  (as diagnosed in the model from KPP), and the local configuration of
  the sigma levels is contoured in gray, showing that the vertical
  resolution of the model increases significantly towards the bottom.}
\label{fig:CC_topow}
\end{figure}

The result is denoted by angle brackets, $\langle \cdot \rangle$, and
we show cross-sections in vertical height $z$ and horizontal distance
across the mean slope $\eta$ (increasing toward deeper water), \eg
$\la \overline{\mathcal{T}}\ra(\eta,z)$ and $\la \overline{w}
\ra(\eta,z)$ in Fig.~\ref{fig:CC_topow}.  Topostrophy is positive for
hundreds of meters above the slope, with maximum values at $\sim$1200
m depth, and the averaged $\la \overline{w}\ra$ profile shows
predominance of strong downwelling within the lowest $\sim$80 m with
weaker upwelling above and even weaker downwelling further into the
interior.

Topostrophy and vertical velocities near the bottom have been also
examined in a solution similar to ALB but forced with tides
\citep{capo2021,capo2023}, with very similar results in the
distribution of positive $\overline{\mathcal{T}}$ and predominantly
negative $\overline{w}_b$, with slightly increased values due to tidal
effects on bottom velocities.  Further ROMS simulations in the
Atlantic and Pacific Oceans examined in \cite{schu2024} --- with
different topographic smoothing, horizontal resolution, and time
averaging over several time scales --- show the same overall
predominance of prograde flows, with positive topostrophy and negative
vertical velocity near the abyssal slopes.  Their work includes
diagnostics from a multitude of current-meter measurements that
  support the hypothesis of prevalent, widespread prograde currents
  and downhill flows.

\section{Mean momentum}
\label{sec:Mbal}
To investigate the various forces sustaining the topostrophic flow, in
this section we analyze the momentum balance in our solution, with a
focus on the Cartagena slope. Then we explore the eddy-driven nature
of the widespread mean prograde flow by decomposing the advective
forcing term into contributions from the mean flow, transient eddies,
and standing eddies.  Finally, an analogous decomposition of the
time-averaged kinetic energy ($KE$) highlights the prominence of eddies
as a principal characteristic.

\subsection{Along-isobath momentum balance and bottom drag}
\label{sec:isobath}
The time-averaged momentum balance in the along-isobath direction is
\begin{equation}
  \widehat{\mathbf{s}}\cdot\pd_t\overline{\bu} \ \approx\ 0 \ =\ \
\underbrace{- \ \widehat{\mathbf{s}}\cdot\overline{ (\vec{u}\cdot\vec{\nabla}) \bu}}_{ADV} \ \
\underbrace{-\ \widehat{\mathbf{s}}\cdot\nabla\overline{\phi}}_{PGF} \ \
\underbrace{-\ f \widehat{\mathbf{s}}\cdot\widehat{\vec{z}} \times \overline{\bu}}_{COR} \ \
\underbrace{+ \ \widehat{\mathbf{s}}\cdot\overline{\mathbf{F}}}_{VMIX} \ .
\label{Ubal}
\end{equation}
$VMIX$ is the parameterized vertical mixing in ROMS described in
Sec.~\ref{sec:model}.  The other right-side tendency terms are
advection of along-isobath momentum ($ADV$) and along-isobath
pressure-gradient ($PGF$) and Coriolis ($COR$) forces.  This balance
is diagnosed on-line and then averaged in time and along-isobath.
Additional terms in the discretized ROMS momentum equations are
negligible in our time averaged momentum balance (\ref{Ubal}); these
include numerical dissipation by the horizontal advection scheme and a
rectification that results from the coupling of the 2D-3D
barotropic-baroclinic, split-explicit time-stepping scheme
\citep{shmc2005}.

\begin{figure}[ht!]  
\centering
{\includegraphics[scale=0.33]{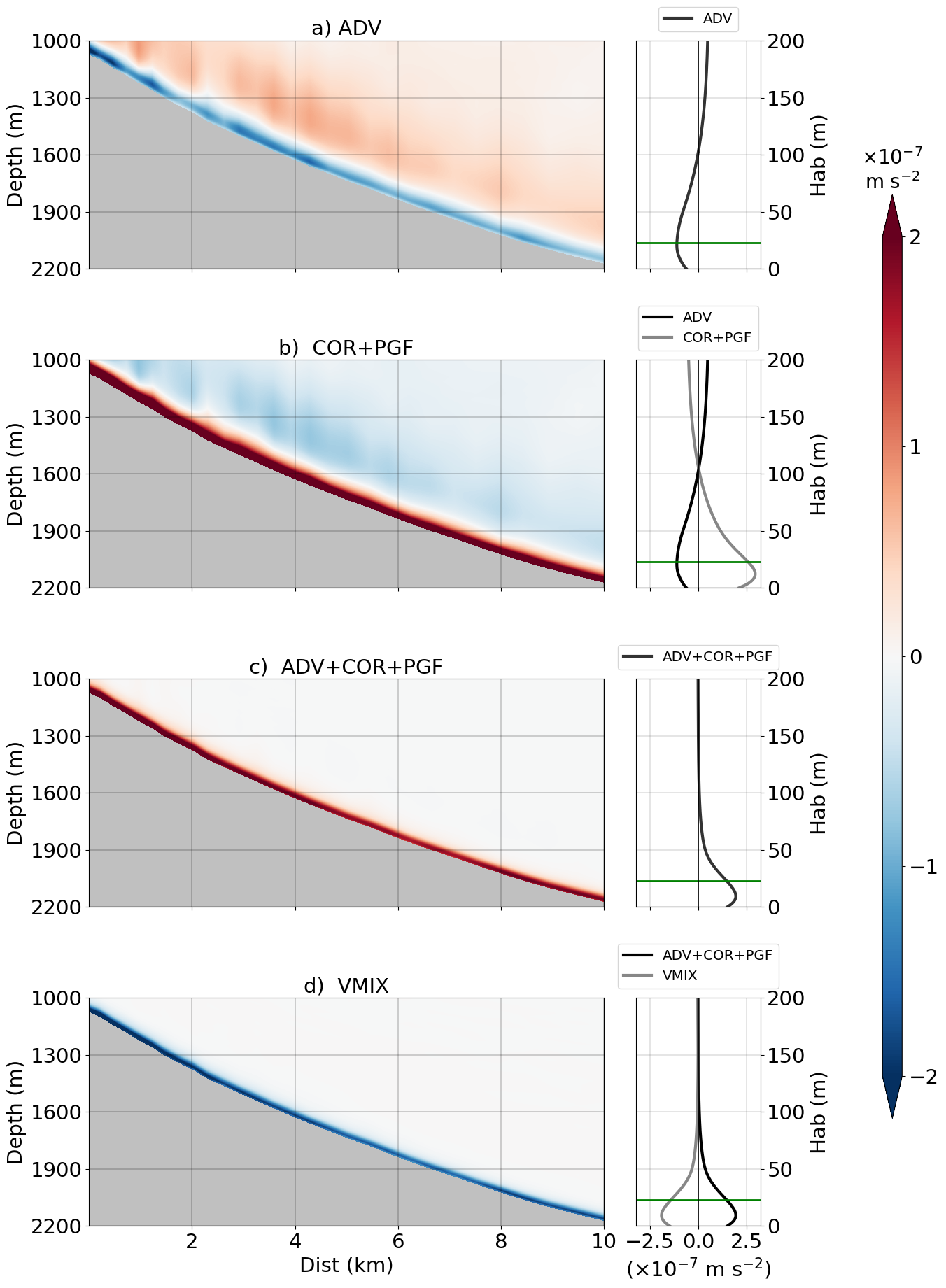}}
\caption{Left panels show time- and along-isobath-averaged
  cross-sections on the Cartagena slope for the along-isobath momentum
  balance (\ref{Ubal}) in the ALB simulation: (a) $\la ADV \ra$, (b)
  $\la PGF \,+\, COR \ra$, (c) $\la ADV \, +\, PGF \,+\, COR \ra$, and
  (d) $\la VMIX \ra$. The 1D profiles on the right panels show the
  corresponding area-averaged terms for the lower 200 m, as a function
  of height above the bottom ($Hab \,=\, h + z$).  While in the
  interior (100 m and above) the balance is between advection,
  Coriolis, and pressure-gradient, in the lower tens of meters
  (including the averaged BBL delimited by the green line), the
  accelerating combination of Coriolis, advection, and
  pressure-gradient is compensated by the retarding effect of bottom
  drag expressed in the vertical mixing term.}
\label{fig:CC_Ubal_slope}
\end{figure}

Figure~\ref{fig:CC_Ubal_slope} shows the momentum balance terms in
various combinations.  On the left panels, geostrophic balance is
hidden in $COR + PGF$, leaving behind a residual that is mostly
balanced by $ADV$, especially in the interior away from the
slope.  $VMIX$ is small except within the BBL, where it reflects the
retardation of the topostrophic current by bottom drag stress and its
vertical redistribution by momentum mixing.  Because the force balances
exhibit continuity along the slope, the slope-averaged profiles shown
in the right panels as a function of height above the bottom are
representative of the balances in the BBL and adjacent interior.
These balances broadly support the ideas of conservative dynamics in
the interior and an Ekman-like balance in the BBL (but see
Sec.~\ref{sec:1D}).

\subsection{Eddy-driven topostrophic flow}
To explore the influence of the eddies in driving the topostrophic
flow, we now decompose the advective term $ADV$ into the contributions
of the mean flow and eddy components.  Given that the flow structure
near the slope in this region is highly modulated by topography
(Fig. \ref{fig:Cartagena}), we chose to make a decomposition of the
flow as follows: a flow component $ u_i(\vec{x},t)$ at a given
location $\vec{x}$ and time $t$ can be written as
\begin{equation}
u_i(\vec{x},t) = \la \overline{u_i} \ra + \widetilde{u_i} + u_{i}^\prime \ ,
\label{flowdec}
\end{equation}
where $\la \overline{u_i} \ra$ is the mean flow, defined here as the
time- and along-isobath average; $\widetilde{u_i} = \overline{u_i} -
\la \overline{u_i} \ra$ is the standing (or stationary) eddies, and $
u_i\prime = u_i - \overline{u_i}$ is the transient eddies.  The
standing eddy component accounts for the persistent circulation
patterns that, in this particular case, are mainly caused by the
interaction of the flow with topographic irregularities, such as
canyons and bumps; the transient eddy component includes temporary
eddies that are typically generated by instabilities in the flow
field, such as ageostrophic shear instabilities and geostrophic
barotropic or baroclinic instabilities.

Using (\ref{flowdec}), the time-averaged advective term in the momentum equation can be written as
\begin{equation}
\overline{(\vec{u}\cdot\vec{\nabla}) \textbf{u}} = \underbrace{  (\la\overline{\vec{u}}\ra \cdot\vec{\nabla})\la\textbf{u}\ra}_{Mean}+ \underbrace{ (\la\overline{\vec{u}}\ra \cdot\vec{\nabla} )\widetilde{\textbf{u}} +(\widetilde{\vec{u}}\cdot\vec{\nabla}) \la\textbf{u}\ra  + (\widetilde{\vec{u}}\cdot\vec{\nabla})  \widetilde{\textbf{u}}}_{Standing \; eddies} + 
\underbrace{\overline{(\vec{u}^\prime\cdot\vec{\nabla})\textbf{u}^\prime}}_{\substack{Transient \\ eddies}} \ .
\label{advec}
\end{equation}
The first term on the right side is the contribution of the mean flow,
the next three terms represent the contribution of the standing
eddies, and the last term represents the advection by transient
eddies.  By projecting (\ref{advec}) onto the along-isobath direction
$\widehat{\mathbf{s}}$ we obtain the corresponding mean, standing
eddy, and transient eddy contributions to the $ADV$ term in
(\ref{Ubal}).

Using an analogous decomposition, the total time-averaged $KE$ is
\begin{equation}
\frac{1}{2} \overline{\vec{u}^2} = \underbrace{\frac{1}{2}\la\overline{\vec{u}}\ra ^2}_{Mean}+ \underbrace{ \widetilde{\vec{u}}\cdot\la\overline{\vec{u}}\ra +\frac{1}{2}\widetilde{\vec{u}}^2}_{\substack{Standing \\eddies}} + 
\underbrace{\frac{1}{2}\overline{\vec{u}^{\prime \,2}}}_{\substack{Transient\\eddies}} \ \ ,
\label{ke}
\end{equation}
where the first term on the right side is the $KE$ of the mean flow;
the next two terms constitute the eddy kinetic energy ($EKE$) stored
in standing eddies; and the last term accounts for the $EKE$ of the
transient eddy flow.

\begin{figure}[ht!]  
\centering
{\includegraphics[scale=0.5]{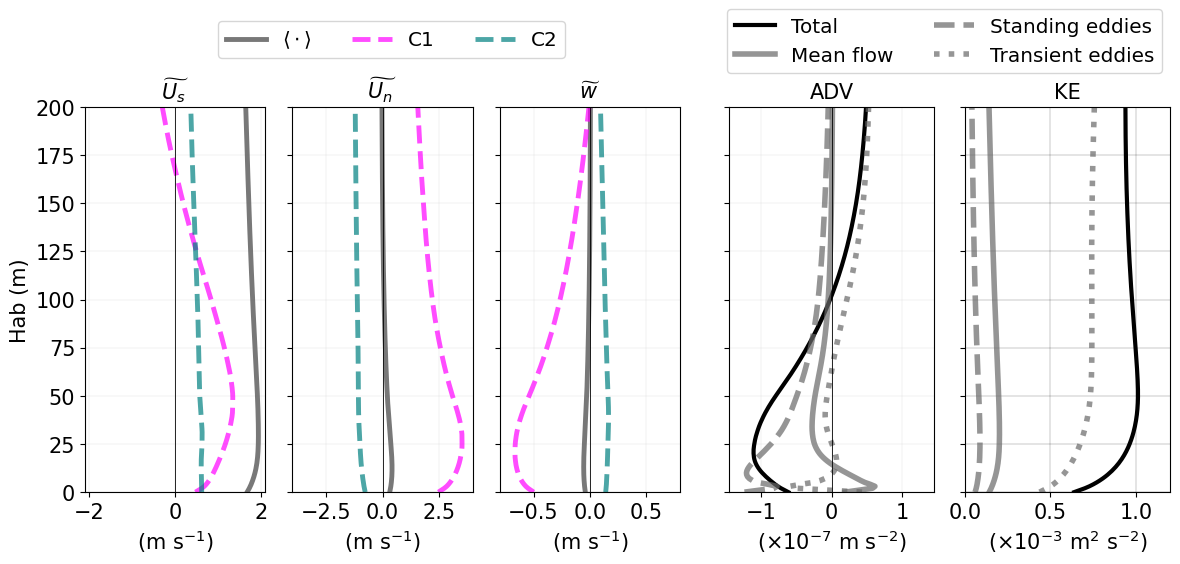}}
\caption{The left three panels show time-averaged profiles of the the
  along-slope ($\widetilde{U_s}$), cross-slope ($\widetilde{U_n}$),
  and vertical ($\widetilde{w}$) velocities, both for the
  area-averaged flow in the Cartagena region (dark gray) and for two
  specific canyon locations shown in Fig. \ref{fig:Cartagena}, C1 and
  C2 (pink and blue shadded lines, respectively).  While the flow is
  generally topostrophic ($\widetilde{U_s}>0$, it is oriented uphill
  in the upstream side of a bump (C2, with $\widetilde{U_n}<0$ and
  $\widetilde{w}>0$)and downhill in the downstream side (C1, with
  $\widetilde{U_n}>0$ and $\widetilde{w}<0$).  The right two panels
  show the time- and area-averaged profiles of the decomposed
  advective term in the momentum balance ($ADV$) and Kinetic energy
  ($KE$) separated into the total (black), mean (gray), standing eddy
  (dashed gray), and transient eddy (dotted gray) components.  In the
  lower 100 m ADV is dominated by standing eddies, and by the
  transient eddy component above, while transient eddies are also the
  principal source of kinetic energy at all levels.}
\label{fig:canyon_ADV_KE}
\end{figure}

The advection and $KE$ results of this flow decomposition are
summarized in Fig.~\ref{fig:canyon_ADV_KE}.  To highlight the relevant
influence of topographic interaction in driving the abyssal sloping
currents and to provide an idea of its vertical extent, in the first
three panels we compare the time-averaged standing eddy component of
the along-isobath $\widetilde{U_s}$, the cross-isobath
$\widetilde{U_n}$, and the vertical $\widetilde{w}$ components of the
flow for three cases: the area-averaged flow (dark gray line), and at
two representative locations at both sides of a bump, C1 and C2 (pink
and blue dashed lines, respectively; see Fig. \ref{fig:Cartagena}). In
all situations, the along-isobath component is topostrophic, while the
cross-isobath and vertical components are in opposite directions at
both sides of the bump, and locally much more intense than the
widespread downhill flow.

Figure \ref{fig:canyon_ADV_KE} also shows that standing eddies
dominate the mean advection of momentum in the lower $\sim$80 m, while
transient eddies are the dominant contribution above.  Transient
eddies dominate the mean $KE$ at all depths near the slope.  These
results provide evidence of the eddy-driven nature of the topostrophic
flow in our abyssal sloping currents.  However, a more complete
analysis of abyssal eddy dynamics, perhaps as a function of spatial
scale, is needed to explain the widespread topostrophic outcome.  Because
the statistical stationarity of the solution is imperative in such
analysis, we choose to defer this until we can achieve a
higher-resolution simulation with a longer time-averaging period.  In
lieu thereof, the along-isobath averaged momentum balance that we
present here is broadly in line with previous studies that show the
turbulent character of the forces driving the mean along-isobath flow
within the BBL \citep{mole2015,chen2021,stew2024}.  \cite{gara2019}
and \cite{spin2021} present observational evidence for the
manner in which topographically localized turbulence is generated
through submesoscale overturning instabilities in the near-boundary
flow (\eg symmetric and centrifugal instabilities; \citet{weth2020})
that arise in the measured topostrophic and downhill mean slope flow.
In Sec.~\ref{sec:intro} some further comments are made about present
theoretical ideas about the cause of topostrophy.  Such arguments are
clearest in barotropic or isopycnal-layer models, whereas the ocean
(and ROMS) is continuously stratified.  Some attempts have been made
to extend the theory to three dimensions \citep{meho1999,vena2012},
but not yet with much generality.

From a horizontal momentum-balance perspective, an important
distinction is what averaging framework is chosen to encompass a
region of complex topography.  Beyond time averaging, one can choose
between a mean horizontal direction (\eg an average $\la
\widehat{\mathbf{s}} \ra$ over the broad slope area) or a local-along
slope direction ($\widehat{\mathbf{s}}$, as here).  \citet{stew2024}
analyzes idealized simulations of corrugated slope flows and finds
that the near-bottom topostrophy-accelerating force is $PGF$ with a
mean along-shore average, whereas the same simulations show $ADV$
acceleration using an along-isobath average, consistent with our
finding here.  In more realistic simulations of the California
Undercurrent, another topostrophic slope flow, averaging more broadly
in the along-isobath direction implicates a $PGF$ acceleration cause
\citep{mole2015,chen2021}.  Because $ADV$ and $PGF$ are closely
related quantities in incompressible fluid dynamics that manifest
turbulent behaviors, there is no contradiction in these different
diagnostic perspectives.

The choice of an averaged direction for the momentum balance might be
simple for an unidirectional flow past an isolated bathymetric
feature, but it can become ambiguous for complex currents and
multi-scale topography as in these simulations.  This led us to choose
the local $\widehat{\mathbf{s}}$.  Another important issue is whether
the direction should be spatially smoothed; \ie on what spatial scales
is topostrophy more likely to occur?  We have not yet investigated
this, but notice that the general pattern of $\overline{\mathcal{T}}$
is quite similar in the WMED and ALB simulations with different grid
resolutions by a factor of three (Fig.~\ref{fig:WM_To_wb}).

\section{Near-slope buoyancy balance and diapycnal mixing}
\label{sec:Bbal}
The widespread negative $\overline{w}_b$ fields observed in the
Western Mediterranean simulations, as well as in the Atlantic and
Pacific solutions presented in \citet{schu2024}, do not match the
behavior expected by the 1D BBL model by \citet{garr1990}, which
predicts up-slope mean current at the bottom ($\overline{w}_b$ $>$ 0)
and negative topostrophy. In this section we analyze the mean BBL
buoyancy balance (\ref{mean-bal}) for the ALB solution to show the
vertical structure near the slope. The contradiction between our
results and the predictions in \citet{garr1990} is further examined in
an alternative 1D BBL model in Sec.~\ref{sec:1D}.

The buoyancy balance terms in (\ref{mean-bal}) are obtained in ROMS
from the online-diagnosed advection and mixing terms in the balances
for potential temperature $T$ and salinity $S$.  They are combined
into buoyancy fluxes with a local linearization of the equation of
state (EOS),
\begin{equation}
\delta b \,=\, -\,\dfrac{g}{\rho_0}\, (\,-\,\alpha \, \delta T 
\,+\,\beta \, \delta S\,) \,,
\label{TSb}
\end{equation}
where $\delta$ denotes the local differential change in the balance
terms and $\alpha = \dfrac{\partial \rho}{\partial T}$, and $\beta =
\dfrac{\partial \rho}{\partial S}$ are the local density expansion
coefficients.  For example, for advection,
\begin{equation}
\overline{\vec{u} \cdot \vec{\nabla}b} \,=\, -\, \dfrac{g}{\rho_0}\,
(\,-\,\alpha \overline{\vec{u} \cdot \vec{\nabla}T} \,+\, \beta
  \overline{\vec{u} \cdot \vec{\nabla}S} \,) \,,
\end{equation}
with an analogous combination in $\vec{\nabla} \cdot
\overline{\vec{\mathcal{B}}}$ for the mean $T$ and $S$ mixing terms.
ROMS uses a nonlinear EOS that corresponds to the UNESCO formulation
as derived by \cite{jamc1995} that computes \textit{in situ} density
as a function of potential temperature, salinity and pressure, the
latter approximated as a function of depth.  This EOS is then
``stiffened'' by a split into an adiabatic and a compressible part to
reduce pressure-gradient errors associated with nonlinear
compressibility effect \citep{shmc2011}.

The method of analyzing in ROMS the total buoyancy mixing term, $-\,
\vec{\nabla} \cdot \overline{\vec{\mathcal{B}}}$, due both to the
explicit subgrid-scale parameterization and to model discretization
errors, is presented in \cite{gula2024}.

The occurrence of a mean downhill slope flow gives an expectation that
the advective tendency in (\ref{mean-bal}) will be positive, \ie
acting to increase the buoyancy.  Of course, advection is three
dimensional and time-varying, including by the mean current and by the
standing and transient eddies. Asuming a nearly stationary state with 
negligible net buoyancy tendency, if the advective tendency is positive,
then the necessary balancing mixing tendency will be negative,
acting to decrease the buoyancy though a divergent $\vec{\mathcal{B}}$
flux.  Because $\widetilde{\omega}$ in (\ref{omega}) has the same sign
as the mixing tendency term, this is a wide layer with diapycnal
downwelling along the slope, contrary to the arguments in
\citet{ferr2016} \etc ~This difference is explained through the use of
an idealized boundary layer model in Sec.~\ref{sec:1D}.

\begin{figure}[ht!]  
\centering
{\includegraphics[scale=0.35]{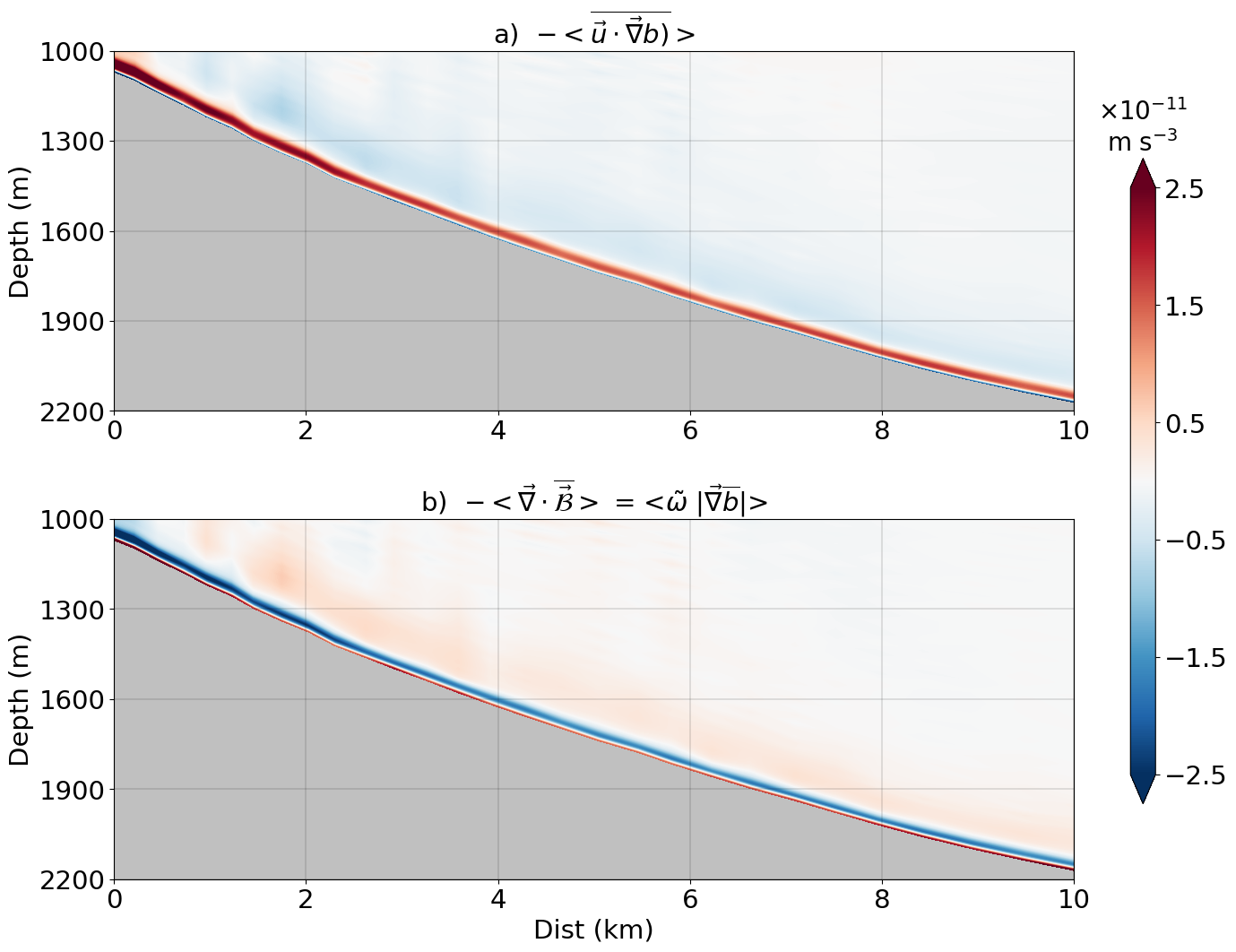}}
\caption{Time- and along-isobath-averaged cross-sections on the
  Cartagena slope for the buoyancy balance (\ref{mean-bal}) in the ALB
  simulation: (a) advection and (b) mixing.}
\label{fig:CC_buoyancy_slope}
\end{figure}

\begin{figure}[ht!]  
\centering
{\includegraphics[scale=0.34]{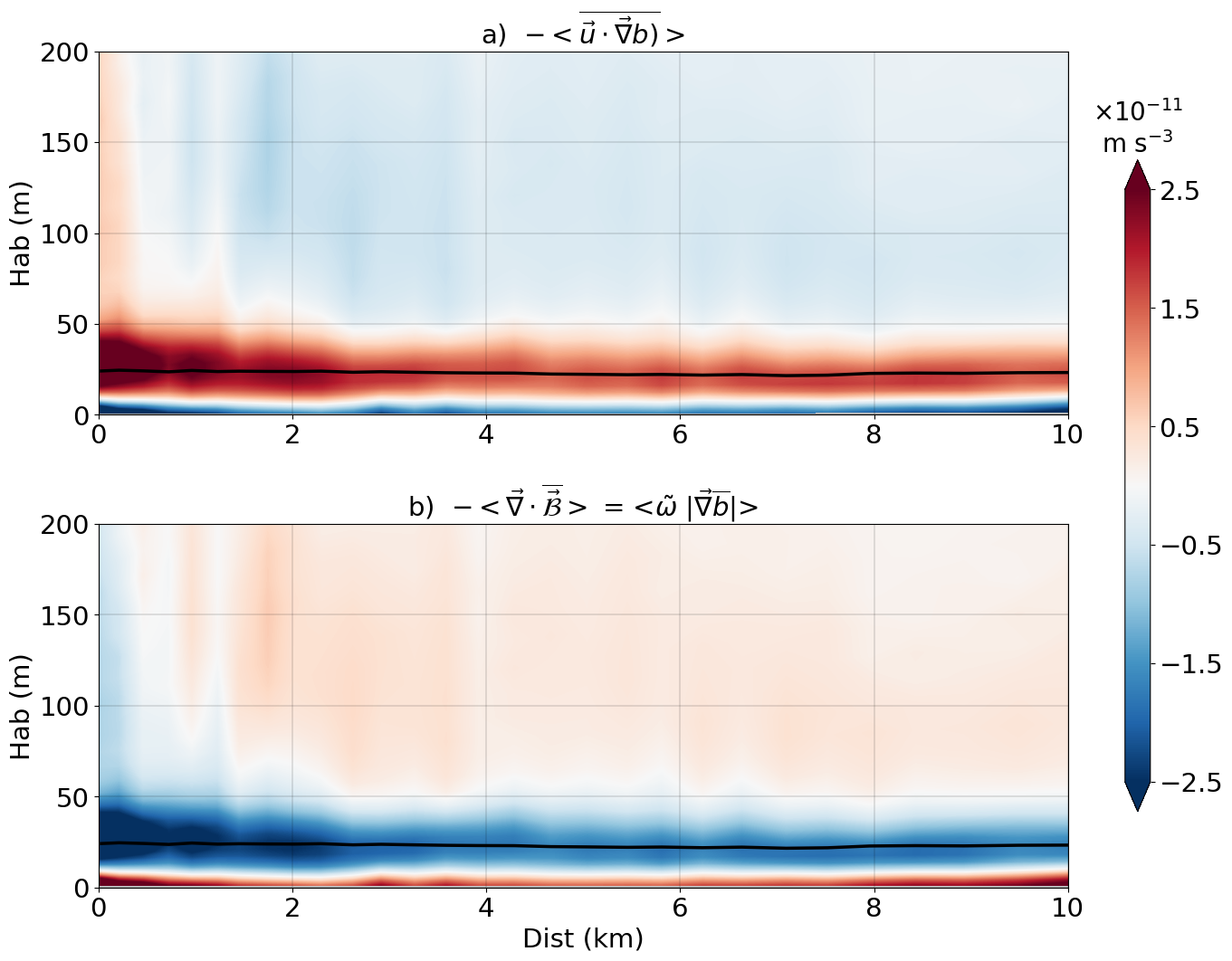}}
\caption{A zoom of the fields in Fig.~\ref{fig:CC_buoyancy_slope} in
  coordinates of height above the bottom $Hab \,=\, z+h$ and
  cross-slope horizontal distance $\eta$.  The mean boundary-layer
  height (as diagnosed in the model from KPP) is the black thick line.}
\label{fig:CC_buoyancy_hab}
\end{figure}

Figure~\ref{fig:CC_buoyancy_slope} shows the time- and
along-isobath-averaged buoyancy balance for the Cartagena region, and a
detail of the lowest 100 m over the slope is shown in
Figure~\ref{fig:CC_buoyancy_hab}.  As anticipated from
$\overline{\mathcal{T}}_b > 0$ and $\overline{w}_b < 0$ in
Sec.~\ref{sec:intro}, the buoyancy advection tendency is positive and
the mixing tendency is negative in the middle and upper part of the
boundary layer, with sign reversals in the adjacent interior.

However, in a thin layer several meters thick right at the slope,
there is a negative advective tendency, contrary to the
$\overline{w}_b < 0$ expectation.  This is a subtle effect of eddy
buoyancy advection as demonstrated below in Fig.~\ref{fig:wpbp} and
further explained in Sec.~\ref{sec:1D}.  The expanded scale in
the BBL close to the slope in Fig.~\ref{fig:CC_buoyancy_hab} shows
that the vertical sign reversals occurs well inside the BBL,
accompanied by a further sign change higher in the interior.  Here we
are using a transformed vertical coordinate, the height above the
bottom, $Hab \,=\, z+h$.

The positive sign of $\widetilde{\omega}$ (\ref{omega}) in the outer
part of the BBL and above means diapycnal upwelling.  Its sign
reversal is consistent with the sign reversal of the cross-isobath
velocity profile in the 1D topostrophic boundary layer model (\ie
$V(Z)$ in Fig.~\ref{fig:sketch} and $V'(Z)$ in Fig.~\ref{fig:pbUV}
below).  Thus, our results are consistent with the idea that much,
maybe most, of the mixing-induced abyssal upelling in the global
thermohaline circulation may be happening near the major topographic
slopes, even if not is confined to a thin bottom boundary layer 
immediately above the slope.

While the tracer mixing in our simulations is prescribed by the
turbulence parameterization (with the vertical eddy diffusivity
$\kappa_v$ derived from the KPP scheme; Sec.~\ref{sec:model}), in
practice mixing of temperature and salinity in the model can
additionally arise from implicit contributions from the horizontal and
vertical advection schemes as well as other discretization errors, in
particular where currents and $T$, $S$ gradients are large.  The
resulting non-advective $T,\, S$ fluxes will result in an
``effective’’ diffusivity, $\kappa_{eff}$, that can differ from
  the prescribed vertical diffusivity $\kappa_v$ \citep{gula2024} and
that is not confined only to a thin bottom boundary layer immediately
above the slope.  In the context of abyssal slope currents,
bottom-intensified mixing due to breaking inertia-gravity waves and
topographic wakes is an important phenomenon (Sec.~\ref{sec:intro})
even beyond the turbulent mixing in the BBL.  With the horizontal grid
resolution of $dx = 500$ m in the ALB simulation
(Sec.~\ref{sec:model}), we can expect an incomplete representation of
this adjacent interior mixing apart from what occurs in $\kappa_v$ due
to small Richardson number in the model-resolved shear and
stratification \citep{larg1994}.  Nevertheless, we can assess both it
and the effective diapycnal mixing in our solution by evaluating the
prescribed $\kappa_{\nu}$ with the diagnosed $\kappa_{eff}$.  The
diagnostic method for the latter is presented in \cite{gula2024} and
illustrated there with ROMS simulations for the Atlantic Ocean.
\begin{figure}[ht!]  
\centering
{\includegraphics[scale=0.45]{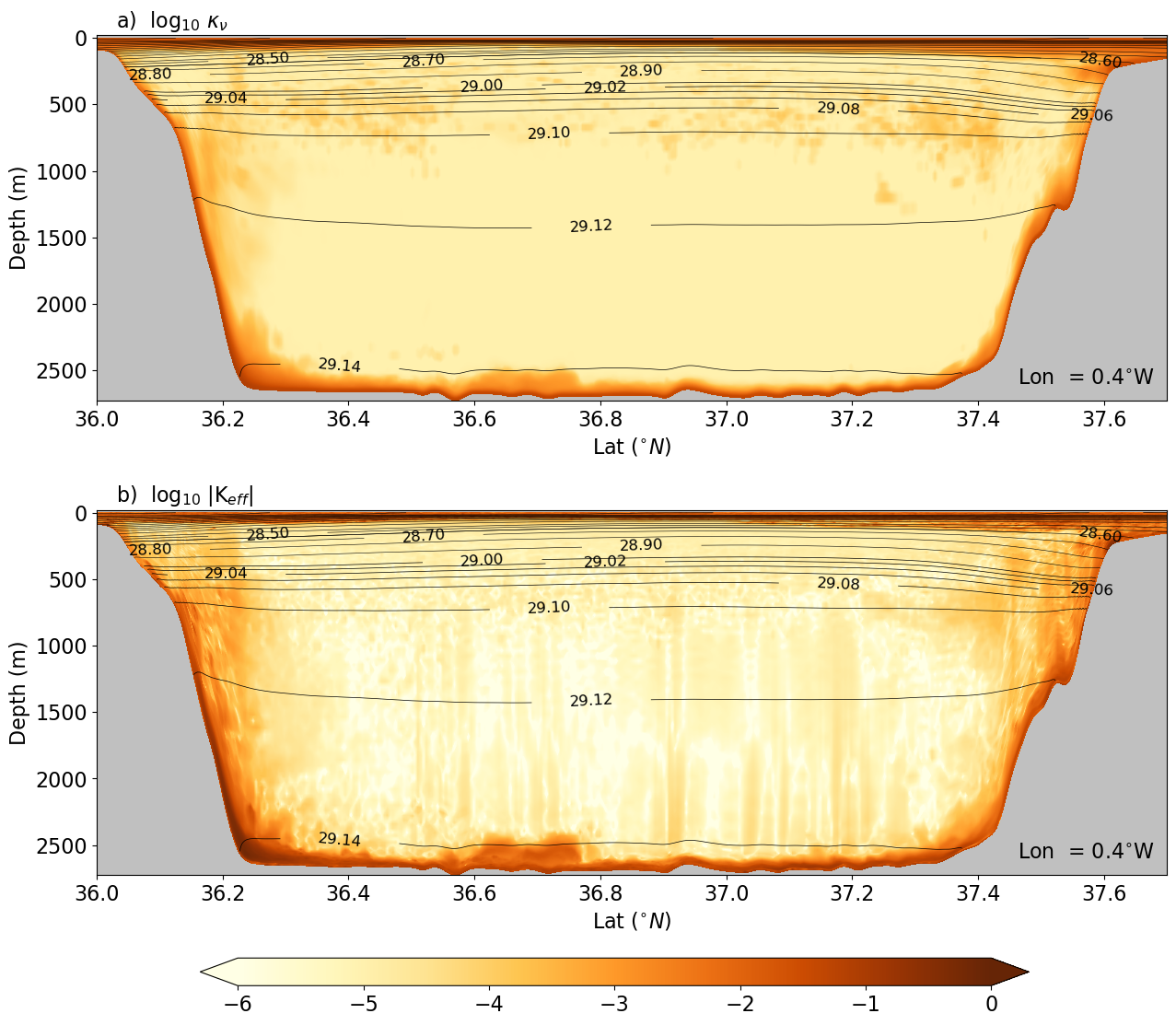}}
\caption{The time-averaged eddy diffusivity $\kappa(y,z)$ [m$^2$
    s$^{-1}$] in the ALB simulation along a section at 0.4$^\circ$W,
  approximately between Oran and the coast near Cartagena (see
  Fig.~\ref{fig:WM_To_wb}b): (a) vertical $\kappa_v$ as diagnosed
  with the KPP parameterization and (b) diapycnal $\kappa_{eff}$ as
  diagnosed from the total diapycnal mixing \citep{gula2024}.  The
  black lines are adiabatic density surfaces.}
\label{fig:K_section}
\end{figure}

By definition, $\kappa_{eff}$ is obtained by projecting
$\mathcal{\vec{B}}$ onto the direction perpendicular to neutral
density surfaces, $\vec{\nabla} b$/$\vert\vec{\nabla} b\vert$, where
$\vec{\nabla} b$ is the adiabatic buoyancy gradient:
\begin{equation}
  \kappa_{eff} = -\, \dfrac{\vec{\mathcal{B}} \cdot
    \vec{\nabla} b}{\vert\vec{\nabla} b\vert^2} \,.
\end{equation}
In the absence of an available online $\kappa_{eff}$ code in our ALB
solution, we make an offline estimation by computing the net
$\overline{\mathcal{\vec{B}}}$ from the diagnosed $T$ and $S$ mixing
fluxes using the locally linearized EOS as in (\ref{TSb}) and the
adiabatic buoyancy gradient via the stiffened nonlinear EOS introduced
above.  The profile in Fig.~\ref{fig:K_section} across the
Mediterranean Sea shows bottom-intensified mixing near the bottom
reaching above the BBL height by hundreds of meters, especially over
the slopes, with magnitudes much larger than in the interior, as well
as elevated values in the surface boundary layer.  We plot
$|\kappa_{eff}|$ because there are occasional small negative values of
$\kappa_{eff}$ due to the uncontrolled sign of model discretization
errors.  These distributions are at least qualitatively consistent
with the abyssal mixing measurements \citep{wate2014}, from
  which interior values of O($10^{-4}$ \,m$^2$s$^{-1}$) are estimated
  in the deep ocean, but increase to O($10^{-2}$m$^2$ s$^{-1}$) near
  topography.  Even if our model's mixing might be quantitatively
inaccurate, we do expect that the simulated structure of the slope
currents, buoyancy balance, and momentum balance (Sec.~\ref{sec:Mbal})
to be robust, partly based on extensive studies of mixing sensitivity
in 1D models (Sec.~\ref{sec:1D}-\ref{sec:discuss}).

\begin{figure}[ht!]  
\centering
{\includegraphics[scale=0.5]{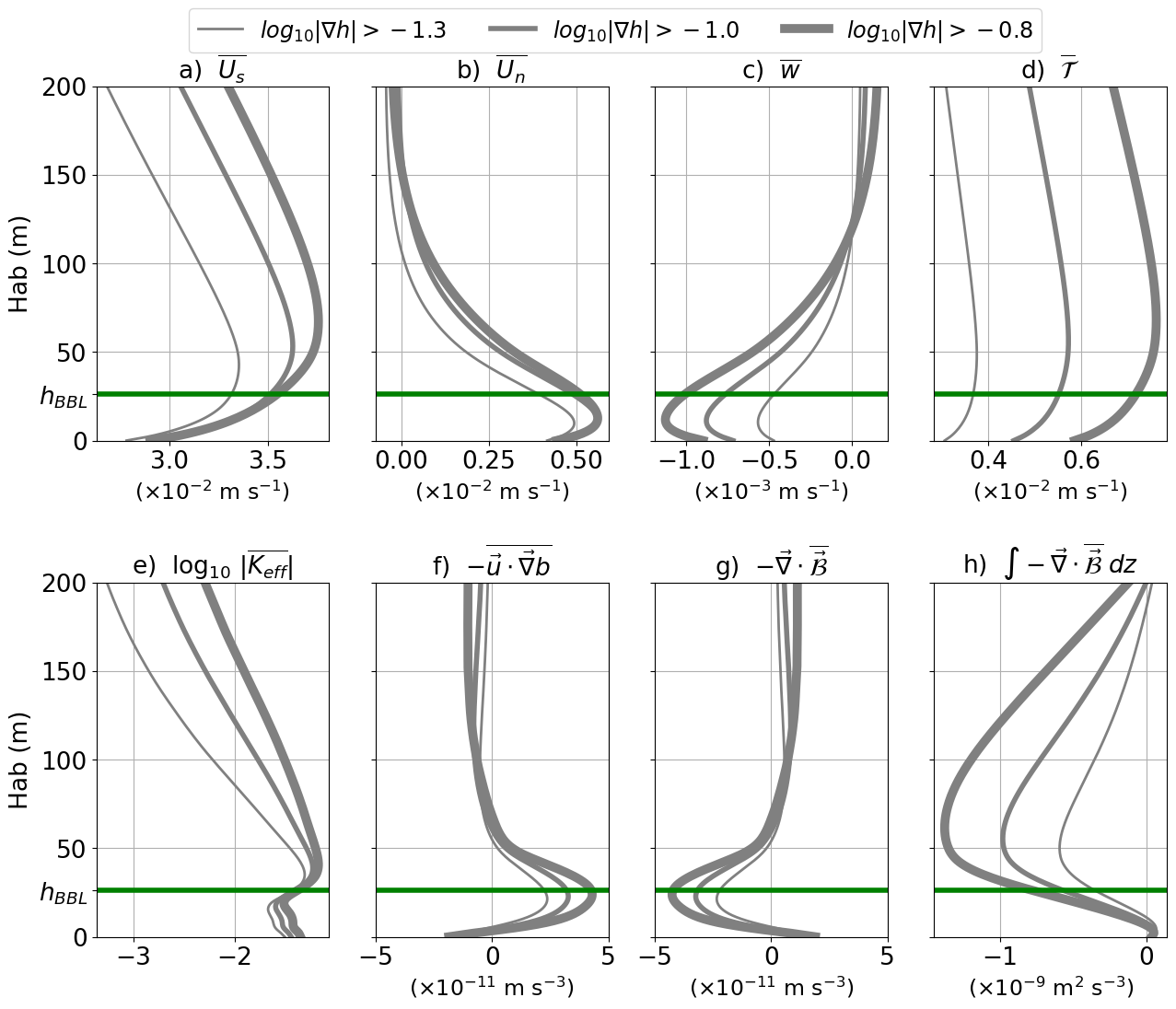}}
\caption{Area- and time-averaged, near-bottom vertical profiles for
  (left to right and top to bottom) (a) horizontal current in the
  along slope (topostrophic) direction; (b) horizontal current in the
  cross slope direction; (c) $\overline{w}$; (d)
  $\overline{\mathcal{T}}$; (e) $log_{10} |\kappa_{eff}|$; (f)
  $-\overline{\vec{u} \cdot \vec{\nabla} b}$; (g) $-\overline{
    \vec{\nabla} \cdot \vec{\mathcal{B}}}$ and (h) the cumulative
  vertical integral of $-\overline{\vec{\nabla} \cdot
    \vec{\mathcal{B}}}$, plotted against height above the bottom
  $Hab$.  The averaging area is the whole of the ALB domain east of
  Gibraltar Strait and with steep slopes, applying different minimum
  thresholds for the slope steepness, $log_{10}|\nabla h|$ (as
  indicated by the line thicknesses).  The green line is the mean BBL
  height.}
\label{fig:area_average}
\end{figure}

To show the typicality of the topostrophic regime, area-averaged
profiles made over steep slopes in the whole of the abyssal Alboran
Sea in the ALB simulation are in Fig.~\ref{fig:area_average}.  These
averages are conditioned on a minimum slope magnitude, $|\nabla h|$;
as this minimum is decreased (from $log_{10}|\nabla h| >-0.8$ to
$log_{10}|\nabla h| >-1.3$), the amplitudes of all the plotted
quantities decrease, but the shapes remain nearly the same.  (Because
these are averages over many locations of quantities that are
nonlinearly related to each other, these averaged profiles should not
be viewed as relevant to any particular place.)  Topostrophic currents
are the norm, reaching well above the BBL; downhill flow occurs
adjacent to the slope; and diapycnal downwelling ($\widetilde{\omega}
> 0$ in (\ref{omega})) occurs in the bulk of the boundary layer.  The
cross-isobath mean flow turns uphill above the boundary layer along
with diapycnal upwelling.  As mentioned above, the mean buoyancy
balance again reverses sign in a thin layer very close to the
slope. The prevalence of net diapycnal downwelling within the
  BBL is reflected by the cumulative integral of the mixing term shown
  in (h), which is negative well into the nearby interior before
  turning positive above.  Finally, the effective diffusivity profiles
  reflect bottom-intensified mixing reaching far above the BBL.

Figure~\ref{fig:wpbp}a shows the decomposition of the advective
tendency term, $-\overline{\vec{u} \cdot \vec{\nabla} b}$, into the
contributions of the time-mean flow, $-\,\overline{\vec{u}} \cdot
\vec{\nabla} \overline{b}$, and the eddies, $-\,\overline{
  \vec{u}^\prime \cdot \vec{\nabla} b^\prime }$. It demonstrates that
the contribution of the mean current (solid gray line) is always
positive near the bottom, increasing the local buoyancy by advecting
less dense water downhill, while the contribution of eddy advection
(dashed gray line) is always negative.  The net buoyancy advection is
thus a competition between these two contributions.  In the bulk of
the boundary layer, the positive mean advection dominates.  However,
the negative eddy advection is the cause of the reversal of the mean
buoyancy balance signs in the thin layer at the slope
(Fig.~\ref{fig:area_average}e).

The source of this negative eddy buoyancy advection is indicated in
Fig.~\ref{fig:wpbp}b, which shows a vertically extensive profile of
negative eddy buoyancy flux, $\overline{w^\prime b^\prime}\,<\, 0$, that
is also the rate of conversion of eddy kinetic energy to eddy
potential energy.  This implicates mean-flow shear instability as the
source of eddy energy followed by the conversion to potential
energy. Furthermore, because the principal gradients of the mean
currents on the broader scale of the abyssal slopes are vertical
(Fig.~\ref{fig:area_average}a,b), it seems likely that it is a
vertical shear instability phenomenon, \ie an ageostrophic process on
a scale larger than boundary-layer turbulence.  The negative sign of
$\overline{w^\prime b^\prime}$ is inconsistent with BBL baroclinic
instability and thus is contrary to the circumstances analyzed in
\citet{call2018} and \citet{drak2022}.

The 1D profiles in Fig.~\ref{fig:wpbp} plotted \vs $Hab$ show the
area-averaged values for the Cartagena domain (solid) and for the
whole Alboran Sea domain in regions with the sea-floor deeper than
800 m.
\begin{figure}[ht!]  
\centering
{\includegraphics[scale=0.6]{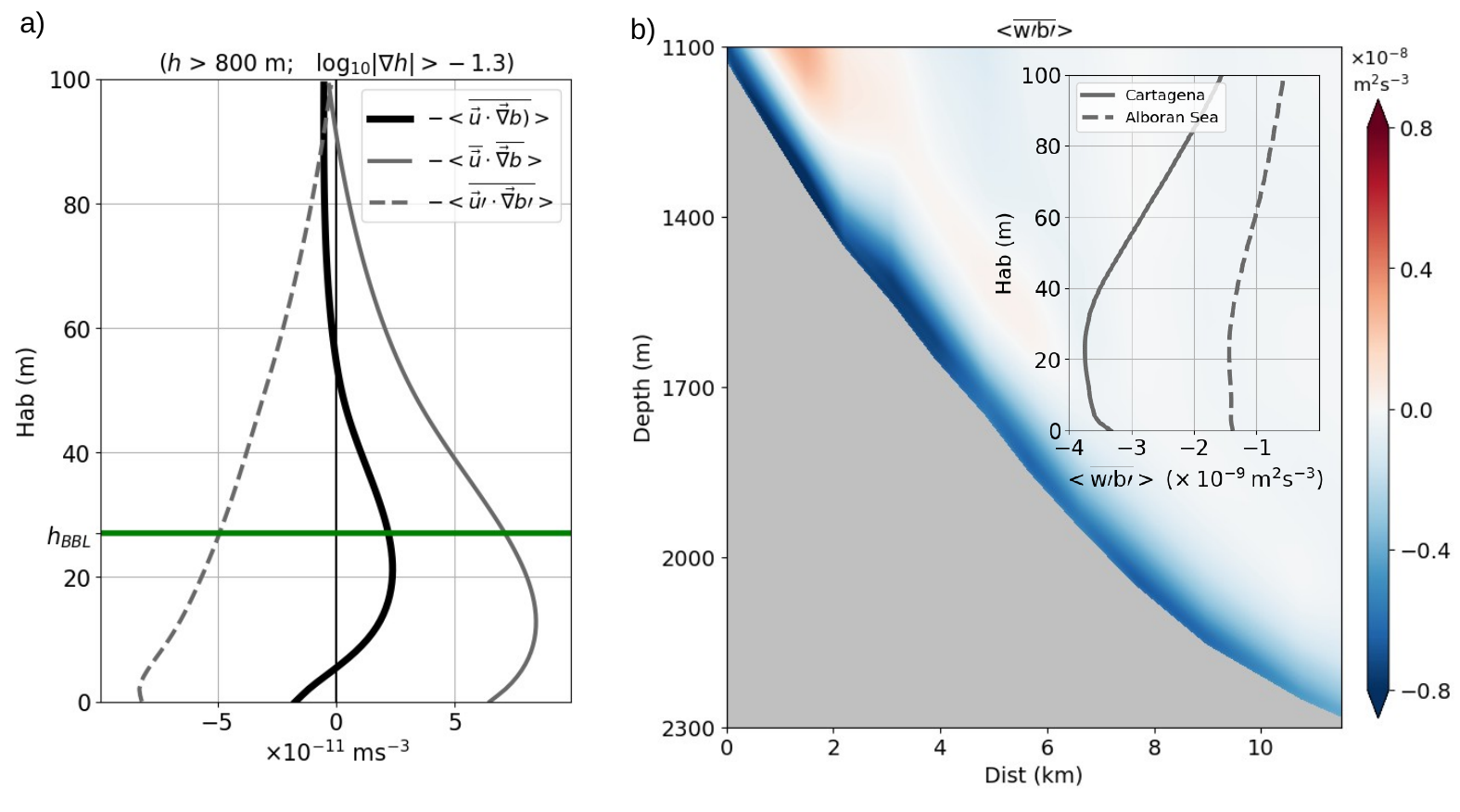}}
\caption{(a) Area- and time-averaged near-bottom vertical profiles for
  $-\,\overline{\vec{u} \cdot \vec{\nabla} b}$ (as shown in
  Fig.~\ref{fig:area_average} for $log_{10}|\nabla h| \,>$ -1.3;
  black); the corresponding input from the time-mean flow
  $-\,\overline{\vec{u}} \cdot \vec{\nabla} \overline{b}$ (solid
  gray); and the eddy contribution $-\,\overline{\vec{u^\prime} \cdot
    \vec{\nabla} b^\prime}$ (dashed gray), plotted against height
  above the bottom $Hab$.  The averaging area is the whole of the ALB
  domain, east of Gibraltar Strait, for $h \ge 800$ m.  The green line
  is the mean BBL height.  (b) Time- and along-isobath-averaged
  cross-sections on the Cartagena slope for the baroclinic term
  $<\overline{w^\prime b^\prime}>$, which entails a conversion from
  eddy available potential energy to eddy kinetic energy through
  baroclinic instabilities if the term is positive. The profiles in
  the inset box correspond to the area-averaged quantities plotted
  against $Hab$ and for $h \ge 800$ m in the same region (Cartagena
  slope; solid) and for the whole ALB domain east of Gibraltar Strait
  (dashed).}
\label{fig:wpbp}
\end{figure}

\section{1D topostrophy-mixing-slope BBL}
\label{sec:1D}
To provide a framework for interpreting the flow structure and
dynamical balances in the preceding sections, we redo the 1D BBL
analysis following \cite{garr1990} but insisting on a topostrophic
interior flow and also including eddy buoyancy transport.  The
existence of the latter effects are demonstrated by the buoyancy and
momentum analyses in Secs.~\ref{sec:Mbal}-\ref{sec:Bbal}, and they are
due to dynamical processes that are not directly associated with the
boundary layer turbulence.

The 1D model is for a steady flow and buoyancy profiles adjacent to a
uniform slope.  In the interior, the solution asymptotes to a uniform
along-isobath horizontal velocity $U_\infty$ with uniform
stratification $b_\infty(z) = N^2z$, both in geostrophic, hydrostatic
balance with the geopotential field $\Phi_\infty(y,z) = \,-\, f
U_\infty y \,+\, N^2z^2/2$.\footnote{
The background stratification is not necessarily a
constant $N$ if $\kappa(z)$ is variable.  Rather than develop a more
general background balance model, we implicitly assume a background
right-side maintenance term in the buoyancy balance, $B_0(z) \,=\, -\,
\pd_z(\kappa N^2)$, to cancel any $N$ evolutionary tendency and to
focus on the boundary layer dynamics.  This is consistent with most
previous 1D BBL analyses, and it might well implicate interior
buoyancy advection and circulation, as in \cite{call2018,drak2022}.}

The model is expressed in rotated $(X,Y,Z)$ coordinates to align with
the sloping bottom at $Z = 0$.  The coordinate transformation is
\begin{equation} 
X = x\,, \quad Y = y \cos[\theta] + z \sin[\theta] \,, \quad
Z = \, -\, y \sin[\theta] + z \cos[\theta] \,,
\label{XYZ}
\end{equation}
where $\theta > 0$ is the angle of the sloping bottom upward to the
north.  In these coordinates,
\begin{align}
  b_\infty &= N^2 (\sin[\theta] Y + \cos[\theta] Z)
  \nonumber \\ \
  \Phi_\infty &= \,-\, f U_\infty\, (\cos[\theta] Y \,-\,
  \sin[\theta] Z ) \,+\, \frac{N^2}{2}\, ( \sin[\theta] Y +
  \cos[\theta] Z)^2 \,.
\label{ff}
\end{align}
The velocity $\vec{u} \,=\, (U,V,W)$ is similarly rotated into this
slope-aligned coordinate system.  For $f > 0$ a topostrophic flow has
$U_\infty < 0$.  Notice that this configuration matches the sketch in
Fig.~\ref{fig:sketch}; \ie $\widehat{\mathbf{s}} \,=\, -\,
\widehat{\mathbf{X}}$ for $f > 0$.

The 1D BBL problem matches this interior solution (with only
along-isobath flow, \ie $V_\infty = W_\infty = 0$) with boundary
conditions at the slope with a no-slip velocity, $\vec{u} = 0$, and no
buoyancy flux, $\kappa \pd_Z b = 0$, at $Z = 0$ for general,
slope-normal eddy viscosity and diffusivity profiles, $\nu(Z)$ and
$\kappa(Z)$.  It seeks solutions that depend only on $Z$ apart from a
uniform Y dependency in $b$ such that $\pd_Y b = N^2 \sin[\theta]$, a
constant, the same as for $\pd_Y b_\infty$.  Thus, the ordinary
differential equation (ODE) system for deviations from the interior
fields (denoted by $b' = b -\,b_\infty$, \etc) is
\begin{align}
-\, f \cos[\theta] V' &= \pd_Z (\nu \pd_Z U') + F' \nonumber \\
f \cos[\theta] U' &= \sin[\theta] b' + \pd_Z (\nu\pd_Z V') \nonumber \\
f \sin[\theta] U' &= -\, \pd_Z \Phi' + \cos[\theta] \, b' \nonumber \\
N^2 \sin[\theta] V' &= \pd_Z(\kappa\pd_Zb') + {B}_e' \,.
\label{1Deqns}
\end{align}
The boundary conditions are
\begin{align}
& U' \,=\, - \,U_\infty\,, \quad V' \,=\, 0 \,, \quad \kappa 
  \pd_Z b' \,=\, -\, N^2\kappa \cos[\theta] \qquad \mathrm{at} \ Z \,=\, 0\,;
  \nonumber \\
  & U', \, V', \, b' \,\rightarrow \, 0 \qquad \mathrm{as} \
  Z \,\rightarrow\, \infty \,.
\label{bc}
\end{align}
These equations are linear in the dependent variables; the only
advective term retained is the cross-isobath buoyancy advection, \ie
$V'$ times the constant $\pd_Y b$.  $W' = 0$ everywhere.  This system
has forcing terms proportional to $F'(Z)$ and ${B}_e'(Z)$ in the
interior of the ODE system and $U_\infty$ and $N^2\cos[\theta]$ in the
boundary conditions, but not all of these are independent as explained
below.

$F'(Z)$ is the boundary layer deviation from the $ADV+PGF+COR \,=\, 0$
interior balance in the along-isobath mean momentum balance
(Sec.~\ref{sec:Mbal}).  It represents the $ADV'+PGF'$ terms associated
with the eddy advection or other process that causes topostrophy but is
otherwise absent from (\ref{1Deqns}); its role is to convey this
effect into the BBL dynamics.  $B_e'(Z)$ represents the eddy buoyancy
advection, $-\, \la \overline{\vec{\bu}'\cdot\vec{\nabla}b'} \ra$ in
Fig~\ref{fig:wpbp}a, also otherwise absent in (\ref{1Deqns}).  Both
$F'$ and $B_e'$ $\rightarrow 0$ as $Z \rightarrow \infty$.

\cite{garr1990} shows how to solve this system when $F' = B_e' = 0$,
partly by recasting it in terms of the cross-isobath streamfunction
$\psi'(Z)$ with $V' \,=\, \pd_Z\psi'$.  By elimination of other
variables, the ODE system for $\psi'$ in a domain $[0,Z_{max}]$ is the
following:
\begin{align}
\pd_Z^2(\nu \pd_z^2 \psi') \,+\, 4\nu_v q^4 \psi' 
&=\, 4\nu q^4 \,\left(\, \kappa_0\mathrm{ctn}[\theta] \,+\,
\frac{1}{N^2 \sin[\theta]}\,\int_0^\infty\, {B}_e'\, dZ \,\right)
\nonumber \\
& \quad  +\, \frac{f\cos[\theta]}{\nu}\, \int_Z^\infty \, F' \, dZ \,-\,
     \frac{\sin[\theta]}{\kappa} \,\int^\infty_Z \, {B}_e'\, dZ
\nonumber \\
\psi' &=\, \pd_Z\psi' \,=\, 0 \quad \mathrm{at} \ Z \,=\, 0 
\nonumber \\
\psi' &\rightarrow\, \kappa_0 \,\mathrm{ctn}[\theta] \,+\,
   \frac{1}{N^2\sin[\theta]} \, \int_0^\infty\, {B}_e'\, dZ \,, \
\pd_Z\psi' \,\rightarrow\, 0
\quad \mathrm{as} \ Z_{max} \,\rightarrow\, \infty \,,
\label{psi-sys}
\end{align}
where $\kappa_0 = \kappa(0)$ and
\begin{equation}
q^4 \,=\, \frac{1}{4}\, \left(\, \frac{f^2 \cos^2[\theta]}{\nu^2} \,+\,
\frac{N^2\sin^2[\theta]}{\nu\kappa} \,\right) \,.
\label{q-def}
\end{equation}

Given a solution for $\psi'(Z)$, we can evaluate the other profiles by
\begin{align}
b'(Z)  &=\, - \, N^2\sin[\theta] \, \int_Z^\infty \, 
\left(\,\psi' \,-\, \kappa_0 \,\mathrm{ctn}[\theta])\, \,-\,
\frac{1}{N^2\sin[\theta]}\,\int_0^\infty\, {B}_e'\,dZ''
\,\right) \,\frac{dZ'}{\kappa(Z')}
\nonumber \\
& \quad -\, \int^\infty_Z\, \left(\, \int^\infty_{Z'} \, {B}_e' \, dZ''
   \,\right) \, \frac{dZ'}{\kappa(Z')}
\nonumber \\
U'(Z)  &=\, \frac{f\cos[\theta]}{\nu}\,\int_Z^\infty \, 
\left(\, (\psi' \,-\, \kappa_0 \,\mathrm{ctn}[\theta] \,-\,
\frac{1}{N^2\sin[\theta]}\,\int_0^\infty\, {B}_e'\, dZ''
\,\right) \,\frac{dZ'}{\nu(Z')}
\nonumber \\
& \quad -\,
\int_Z^\infty \, \left(\int_{Z'}^\infty \, F' \, dZ''\right) \,\frac{dZ'}{\nu(Z')}
\nonumber \\
V'(Z) &=\, \pd_Z \psi' \,,
\label{bUV}
\end{align}
together with the bottom stress, $\mathbf{\tau}\,=\, \nu \pd_Z (U',\,
V')(0)$, and lateral transport anomaly, $\mathbf{T}^\prime \,=\,
\int_0^\infty\, (U',\, V')\,dZ$.  Notice that $U_\infty$ does not
appear explicitly in (\ref{psi-sys})-(\ref{bUV}); thus, the only
forcing terms are $\kappa_0\, \mathrm{ctn}{[\theta]}$, $F'$, and
${B}_e'$, while $U_\infty \,=\, -\, U'(0)$ is implicit.

Our purpose with this model is as a paradigm for explaining the
topostrophic boundary layer structure evident in the ROMS simulations,
rather than as a deliberate fit at any particular location.  In this
spirit, we choose simple boundary layer profile shapes for the forcing
functions, \eg
\begin{equation}
F'(Z) \,=\, F_0 e^{-\ell Z} \,, \qquad 
{B}_e'(Z) \,=\, {B}_{e0} e^{-\mu Z} \,,
\end{equation}
and ``typical'' values for the parameters to illustrate the solution
behavior.  With both analytic solutions (Appendix) and a numerical
solver for (\ref{psi-sys})-(\ref{bUV}), we have extensively explored
different parameter values and different profile shapes for $\nu$,
$\kappa$, $F'$, and ${B}_e'$.  While these choices have
meaningful quantitative consequences for the 1D solutions, their
qualitative shapes are quite robust for positive diffusivities and
mostly negative values for $F'$ and ${B}_e'$, consistent with
the near-boundary mixing, momentum advection, and eddy buoyancy
advection shown in Sec.~\ref{sec:Mbal}-\ref{sec:Bbal}. Therefore, we
restrict our attention here to a representative set of solutions with
constant $\nu$ and $\kappa$; this choice of parameter values is not
delicate with respect to demonstrating the relevant behaviors.  In our
view, the value of the 1D model is as an idealized demonstration of
the influence of topostrophy on the slope BBL --- and as making a
connection with the conventional 1D model of \cite{garr1990} --- rather
than as a quantitative explanation of the simulation results.

The solution procedure is to decompose the solution into superimposable pieces,
\begin{equation}
\psi' \,=\, \psi'_{\kappa_0,B'} \,+\, \psi'_F \,,
\label{part}
\end{equation}
\etc., where $\psi'_{\kappa_0,B'}$ satisfies the boundary-value
problem (BVP) (\ref{psi-sys}) for $F' = 0$ and $\psi'_F$ satisfies the
BVP (\ref{psi-sys}) for all the terms with $\kappa_0$ and
${B}_e'$ set to zero and $F' \ne 0$.  The latter is initially
specified with an arbitrary multiplicative coefficient that is then
renormalized such that
\begin{equation}
U'_F(0) \,=\, - \, (\,U_\infty + U'_{\kappa_0,B'}(0)\,) \,,
\end{equation}
where $U_\infty$ is specified.  This renormalization factor is applied
to all the $\cdot_F$ functions.  The total solution is then evaluated
by summation as in (\ref{part}).

\begin{figure}[htbp]
\begin{center}
{\includegraphics[scale=0.5]{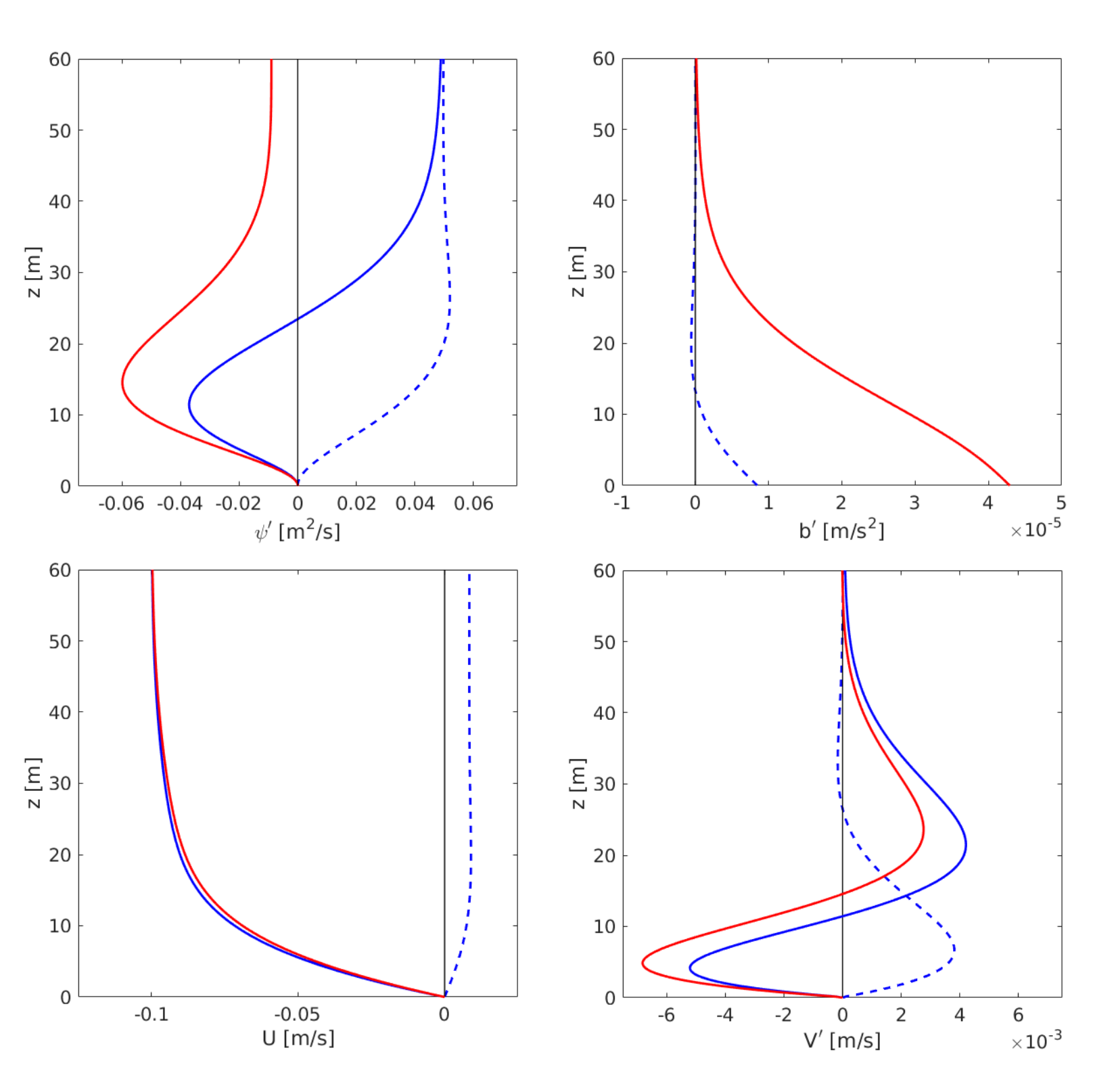}}
\end{center}
\caption{Vertical profiles in the 1D BBL with constant diffusivities:
  from upper-left clockwise, cross-hill streamfunction $\psi'$,
  buoyancy anomaly $b'$, total along-isobath flow $U$, and uphill flow
  $V'$.  The blue dashed lines are for the \textit{mixing BBL}
  ($F'={B}_e'=0$); the blue solid lines are for
  \textit{topostrophic BBL with momentum advection} ($F'\ne 0$,
  ${B}_e' = 0$); and the red solid lines are for the
  \textit{topostrophic BBL with both momentum advection and eddy
    buoyancy advection} ($F',\, {B}_e' \ne 0$).  For $b'$ the
  blue and red solid curves nearly coincide, and for $U$ they are also
  close.  When $F' \ne 0$, $\ell = 0.09$ m$^{-1}$, and $F_0 =
  -\, 5.82 \times 10^{-6}$ or $-\, 4.92 \times 10^{-6}$ m s$^{-2}$ for
  the cases without or with ${B}_{e0} \ne 0$, respectively, to
  assure the value of $U_\infty = - \, 0.1$ m s$^{-1}$.  When
  ${B}_e' \ne 0$, $\mu = 0.05$ m$^{-1}$ and ${B}_{e0} = \,-\,3
  \times 10^{10}$. Other parameters are $f = 10^{-4}$ s$^{-1}$,
  $\kappa = \nu = 0.5 \times 10^{-2}$ m$^2$ s$^{-1}$, $\theta = 0.1$,
  $N^2 = 10^{-6}$ s$^{-2}$, and $q = 0.12$ m$^{-1}$.}
\label{fig:pbUV}
\end{figure}

The 1D BBL solutions with representative parameter values are in
Fig.~\ref{fig:pbUV} for three cases: (1) A \textit{mixing BBL} where
$F'={B}_e'= 0$ and 
\begin{equation}
U_{\infty\,\kappa_0} \,=\,
  \frac{f \kappa \cos[\theta]\,\mathrm{ctn}[\theta]}{\nu q} \,> \, 0 
\label{U_*}
\end{equation}
is fully determined and directed opposite to topostrophic interior
flow \cite[\cf][]{garr1990}; (2) A \textit{topostrophic BBL with
  momentum advection} where ${B}_e'= 0$ and $U_\infty$ has the
specified value by the choice of $F_0$; and (3) A \textit{topostrophic
  BBL with both momentum advection and eddy buoyancy advection} where
both $F'$ and ${B}_e'$ are nonzero.

Figure~\ref{fig:pbUV} shows the primary profiles for the three cases.
As a broad summary statement, the topostrophic boundary layers are
very different from the mixing boundary layer.  $b'(z)$ is positive in
all cases as a reflection of the requirement for the interior
stratification to transition to a no buoyancy flux boundary
condition. The along-isobath flow $U(z)$ monotonically increases from
its no-slip zero value at the slope and has the opposite sign, by
construction when $F' \ne 0$, due to the requirement that it match
with the topostrophic interior value ($U_\infty < 0$ such that
$\overline{\mathcal{T}} > 0$).  The cross-isobath flow $V'(Z)$ also is
in the opposite direction near the boundary with topostrophy; \ie it
is downhill there, consistent with $\overline{w}_b < 0$.  In all cases
$V'$ reverses sign with distance above the boundary; its net lateral
transport $T^{y\,\prime} \,=\, \psi'(\infty)$ is positive in the (1)
mixing and (2) only-topostrophic cases but becomes net negative in
case (3) with ${B}_e' \ne 0$.  These topostrophic BBL solutions in
cases (2) or (3) are the basis for the sketch in
Fig.~\ref{fig:sketch}.

As $U_\infty$ becomes more negative (\ie more topostophic), the
  strength of the near-boundary downhill flow in $V'$ increases; \eg
  for the parameters otherwise as in Fig.~\ref{fig:pbUV} for case (2),
  this $V' < 0$ regime occurs for all $U_\infty < -\, 0.02$ m
  s$^{-1}$.  As $B_{e0}$ becomes more negative, the cross-slope
  transport $T^{y\,\prime}$ becomes more downhill; \eg, for the
  parameters otherwise as in Fig.~\ref{fig:pbUV} for case (3), the
  marginal value for this regime is $B_{e0} = -\, 2.5 \times 10^{-10}$
  m s$^{-3}$. \textit{I.e.},
the essential behaviors of downwhll flow, $V'(0) < 0$, and near-slope
diapycnal downwelling are robust across wide variations in the 1D
model with topostrophy, while the value and sign of the total
cross-slope transport, $\psi'(\infty) \,=\, T^y$, is more variable
depending on the relative importance of ${B}_e'$. 

The signs of these near-slope $V'(Z)$ are consistent with Ekman-layer
thinking in the sense that they are directed to the left of the
along-isobath interior flows $U_\infty$.  However, the sign-reversals
in $V'(Z)$ are inconsistent with the relation between lateral Ekman
transport $\mathbf{T}'$ and the bottom stress $\mathbf{\tau}$ (both
defined following (\ref{bUV})).  In none of these three cases are
either of the lateral Ekman transport relations satisfied.  In
particular, after vertically integrating the $X$-momentum equation in
(\ref{1Deqns}) and using the $\pd_ZU'(0)$ boundary condition in
(\ref{bc}), the cross-isobath transport relation is
\begin{equation}
T^{y\,\prime} \,-\,  \frac{\tau^x}{f\cos[\theta]} \,=\, -\, 
 \frac{1}{f\cos[\theta]} \int_0^\infty\, F'\, dz \,,
\end{equation}
and the nonzero right-side momentum advection breaks the Ekman
relation.  (The along-isobath transport relation has a more
complicated formula, not shown here.)  In the limit of a flat bottom
($\theta \rightarrow 0$) and no topostrophic or eddy buoyancy
advection forcing ($F' = {B}_e' = 0$), these relations revert to
Ekman-layer transport.

\begin{figure}[htbp]
\begin{center}
{\includegraphics[scale=0.5]{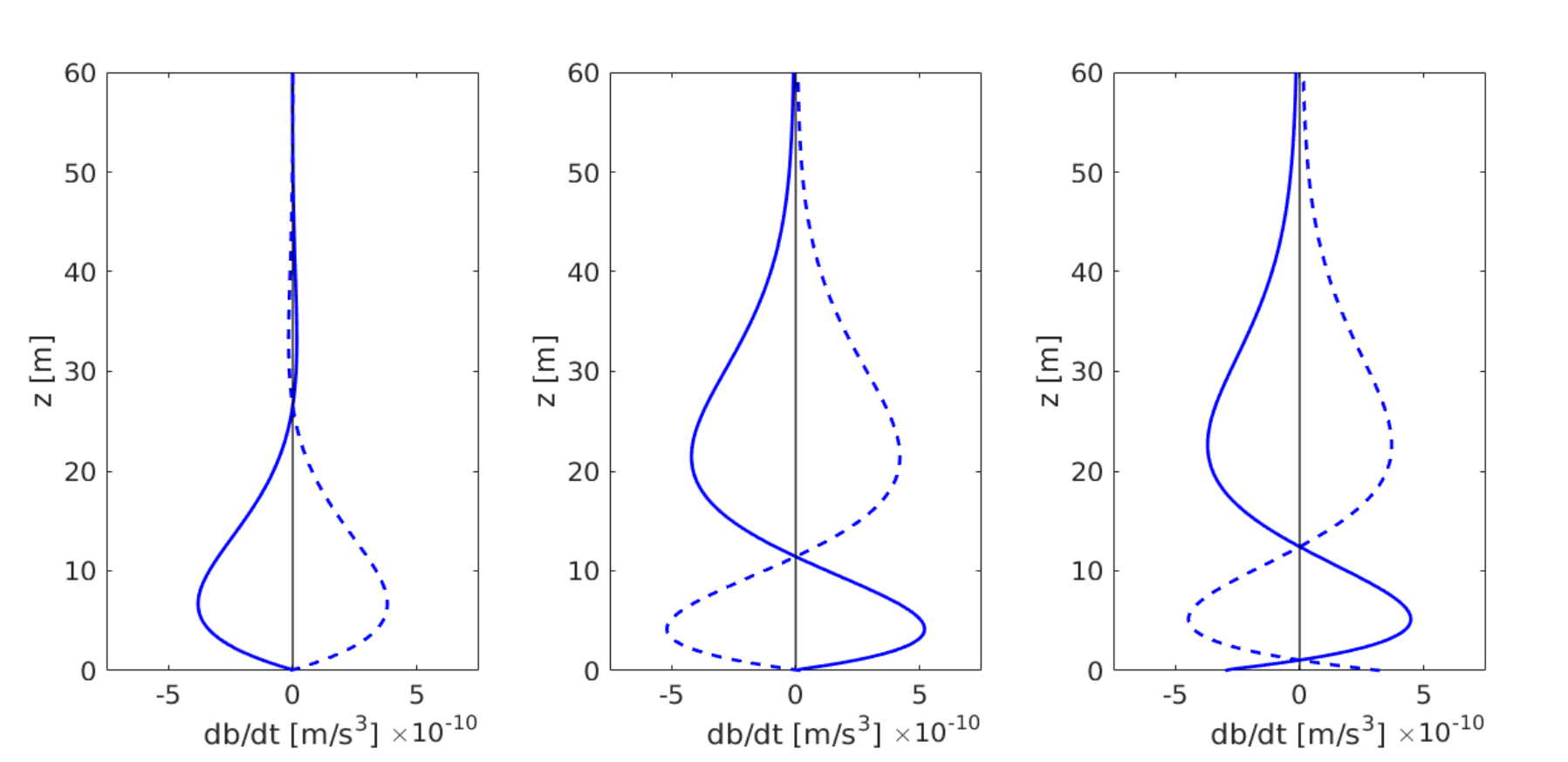}}
\end{center}
\caption{The 1D BBL buoyancy balance terms: mean plus eddy advection
  (solid) and slope-normal diffusion (dashed).  From left to right the
  cases are the mixing BBL ($F'=B_e'=0$), the topostrophic BBL with
  momentum advection ($F'\ne 0$, $B_e' = 0$), and the topostrophic BBL
  with both momentum advection and eddy buoyancy advection ($F',\, B_e'
  \ne 0$).  The parameters are the same as in Fig.~\ref{fig:pbUV}.}
\label{fig:Bbal}
\end{figure}

The eddy buoyancy advection ${B}_e'$ has a moderate quantitative
influence on the topostrophic BBL profiles, slightly so for $b'$ and
$U$ and more so for $\psi'$ and $V'$, but not a qualitatively large
one (Fig.~\ref{fig:pbUV}), except perhaps for the sign of
$T^{y\,\prime}$ that is determined by a competition between buoyancy
mixing and eddy advection.  Where the role of ${B}_e'$ becomes
evident is in comparing cases (2) and (3) for their mean buoyancy
balance in (\ref{1Deqns}) (Fig.~\ref{fig:Bbal}) between net buoyancy
advection, $-\, N^2 \sin[\theta] V' \,+\, {B}_e'$, and turbulent
buoyancy mixing, $\pd_Z(\kappa\pd_Zb')$.  In the mixing-only case (1),
advection tends to decrease the buoyancy as an upslope advection,
while mixing tends to increase it (\ie diapycnal upwelling).  In
contrast, in the topostrophic cases (2) and (3) the lower part of the
boundary layer has advective buoyancy increase and diapycnal
downwelling.  However, in a very thin layer near the slope, case (3)
shows a small amount of diapycnal upwelling due to ${B}_e' \ne
0$ (\cf Fig.~\ref{fig:area_average}e).

The 1D topostrophic BBL model is therefore a useful paradigm for
abyssal slope currents.  Here we have only shown representative
solutions.  The parameter values and profile shapes for $\kappa$,
$\nu$, $F'$ and ${B}_e'$ could be varied to explore their
sensitivities, but it seems to us much more important to look to
further measurements and simulation models like ours to better
determine what the true distributions are for diapycnal mixing,
topostrophic momentum mixing, and eddy buoyancy fluxes near the
abyssal slopes.

The mixing-only 1D BBL model has been very influential in ideas about
abyssal upwelling near slopes (Sec.~\ref{sec:intro}), and the
widespread prevalence of topostrophy is an important corrective to
such arguments, as further discussed in Sec.~\ref{sec:discuss}.
Topostrophy is also an important correction to the ``arrested Ekman
layer'' idea that initial interior flows on slopes evolve toward the
mixing-only steady state with a weak $U_\infty$
\citep[\eg][]{garr1993,brle2010,ruan2019,ruan2021} when there are no
confounding interior dynamics to counter this evolution.  In 3D flows
with sustained interior flows, the tendency toward arrest is weaker
\citep{jaga2023}, in particular in topostrophic situations.

We can contrast this 1D steady, stratified, slope BBL model with the
conventional 1D steady Ekman layer model for unstratified flow above a
flat bottom, both with constant eddy mixing coefficients. The
connection between these simple models is not direct; \ie in the limit
of $\theta \rightarrow 0$ with other parameters held constant, the
former has $(U_{\kappa\,\infty},\, V')\rightarrow \infty$ because
(\ref{1Deqns}-\ref{bc}) imply non-zero stratification and lateral
buoyancy advection (see Appendix formulas).  In other words, the
conventional Ekman BBL model does not encompass stratification, which
is an essential part of the slope BBL models.  To properly span these
two regimes, a more general BBL model is required.

\section{Discussion}
\label{sec:discuss}
This is an era of increasing appreciation of the dynamical importance
of the ocean's abyssal topographic slopes: boundary-layer turbulence,
inertia-gravity wave and vorticity generation, adjacent interior
diapycnal mixing due to breaking waves and wakes, along-isobath
(topostrophic) flows, and eddy fluxes.

The primary contributions in this paper are demonstrating widespread
topostrophy and how this feature and momentum and buoyancy advection
and turbulent mixing in the near-slope region lead to a down-slope
mean flow and extensive diapycnal downwelling, taking a coarse-grained
view beyond small-scale canyons and ridges.  A similar result is
reported in \citet{schu2024} for an analogous simulation in the
Atlantic Ocean.  Relative to the scientific literature, in our view,
this amplifies the importance of eddy-driven topostrophy, and it
provides an important modification to the recent advocacy of
slope-adjacent diapycnal upwelling
\citep{ferr2016,dela2016,mcfe2017,cafe2018,dela2016,holm2018,holm2019,
drak2020,homc2020,drak2022,pete2022, bake2023, pete2023,wynne2024observations}.

A strong rationale for the slope-upwelling view has been the 1D slope
BBL model \citep[\eg][]{garr1990} and its elaborations with various
depictions of the vertical profile of bottom-enhanced mixing.  We show
how this is modified by a topostrophic flow in Sec.~\ref{sec:1D} to
reverse the slope-adjacent flow, but still have diapycnal upwelling in
the outer part of the BBL and nearby interior.  The global
thermohaline circulation requires net diabatic and volumetric
upwelling in the abyss, of course, and because of the locally enhanced
diabatic mixing it seems likely that much if not most of it occurs in
topographic neighborhoods.  However, regional models, such as the ones
analyzed here and in \cite{schu2024}, are not the right tool for
assessing the net global upwelling because they are strongly
influenced by imposed abyssal inflows through their open lateral
boundaries.


\section{Summary and Prospects}
\label{sec:sum}
Abyssal slope currents are significantly topostrophic, due, in a
general way, to turbulent ``mesoscale'' oceanic momentum dynamics.
From this phenomenon follow both the downhill mean flow and the
diapycnal slope downwelling and nearby upwelling that are also
widespread phenomena, supported by both boundary-layer turbulence and
adjacent diapycnal mixing.

This paper and its companion \citep{schu2024} comprise a synthesis and
partial revision of several prevailing ideas about oceanic circulation
and mixing: geostrophic topographic turbulence, slope bottom boundary
layers, near-slope mean-flow eddy instability, and abyssal upwelling
due to bottom-intensified diapycnal mixing.  However,
current-topography interaction in the abyss remains a significant
scientific frontier.  Apart from simple idealized configurations, it
involves demanding computational simulation capabilities for mesoscale
eddies, the bottom kinematic boundary condition, bottom boundary layer
turbulence, topographic vertical vorticity and inertia-gravity
wave generation, and submesoscale wakes --- all in their appropriate
context for a realistic regional circulation.

In our view, further progress lies ahead via improved model
  formulations and finer grid resolutions in both idealized and
  realistic configurations. Further investigations are needed for at
  three particular issues in abyssal slope regions: (1) the spatial
  distribution of diapycnal mixing in the mean buoyancy balance, (2)
  the advective ``eddy-driving'' for topostrophic mean flow and the 
  dynamical conditions under which it is a dominant effect, and
  (3) local currents in canyons and ridges that underlie the generally
  topostrophic slope flow, including tidal effects.

\acknowledgments The authors acknowledge helpful discussions with
Andrew Stewart.  Support for this research was provided by the Office
of Naval Research (ONR grants \#N00014-18-1-2599 P00001 and
\#N00014-23-1-2226) and the National Science Foundation (NSF grant
\#2241754).  J.G. and R.S. gratefully acknowledge support from the
French National Agency for Research (ANR) through the project DEEPER
(ANR-19-CE01-0002-01).  Computing support was provided by the NSF
ACCESS program.

%
%
\datastatement The ALB model simulation used in this work was
performed using the UCLA version of the ROMS model
(https://github.org/CESR-lab/ucla-roms.git).  Configuration files and
an output sample are available at
https://doi.org/10.5281/zenodo.10215065 \citep{capo_ALB500UR_2023}.


%






%



\appendix
\appendixtitle{1D BBL relations}
Here we present 1D BBL solutions in analytic form.  For brevity we do
not include cases with $\mathbf{B}' \ne 0$.  The problem is posed in
Sec.~\ref{sec:1D} and representative solutions are in
Fig.~\ref{fig:pbUV}.  Because of the linearity of (\ref{psi-sys}), we
can superimpose the separate solutions for each forcing; \eg
\begin{equation}
U' \,=\, U'_\kappa + U'_F \,.
\label{superpose}
\end{equation}
$U'_\kappa$ is the solution with $F'= 0$, and $U'_F$ is the solution
with a homogeneous buoyancy boundary condition, $\pd_Zb' = 0$.  We
designate them as the buoyancy-mixing and topostrophic components,
respectively.   For each of these, $U_\infty$ is fully determined.  For
$U'_F$ we will take the inverse view that $F'(Z)$ is chosen such that
$U_\infty$ has the desired topographic value.

The solutions specified below are plotted in Fig.~\ref{fig:pbUV} for
the parameter values listed in the caption.

\vskip -30pt
\subsection{Mixing BBL}
First consider the case with $F' = 0$ (referred to as case (1) in
Sec.~\ref{sec:1D}).  Its solution with constant $\nu$ and $\kappa$
is presented in \cite{garr1990}.  The formulas are
\begin{equation}
  \psi'_\kappa(Z) \,=\, \kappa \, \mathrm{ctn}[\theta] \, \left(\, 1 \,-\, e^{-qz}
  (\sin[qz] + \cos[qz]) \,\right) \,.
\end{equation}
The other profiles are
\begin{align}
  b'_\kappa(Z) &= \frac{N^2\cos[\theta]}{q}\, e^{-qz} \cos[qz]
  \nonumber \\
  U'_\kappa(Z) &= -\,U_* \, e^{-qz} \cos[qz]  \,,
  \nonumber \\
  V'_\kappa(Z) &= 2 q \kappa \, \mathrm{ctn}[\theta] \, e^{-qz} \sin[qz] \,,
\end{align}
where the surface value of the along-isobath velocity anomaly is
$U'_\kappa(0) \,=\, -\, U_*$ with
\begin{equation}
  U_* \,=\,
  \frac{f \kappa \cos[\theta]\,\mathrm{ctn}[\theta]}{\nu q} \,> \, 0 \,.
\label{U*}
\end{equation}
As anticipated, this solution is consistent only with a retrograde
interior along-isobath velocity, \ie the opposite of topostrophy.

\subsection{Topostrophic BBL}
Now consider the solution with $F' \ne 0$ with no terms explicitly
proportional to $\kappa$ in (\ref{psi-sys})-(\ref{bUV}), because they
are accounted for in the mixing BBL solution, but still with a
$\kappa$ dependency in both $q$ and the denominator of the coefficient
for $b'(Z)$.  For the 1D BBL problem, $F'$ is not known \textit{a
  priori} because it is considered to be a product of the 3D interior
topostrophic dynamics.  This product is conveyed into the BBL problem
by requiring that the total $U(Z) \rightarrow U_\infty$ as $Z
\rightarrow \infty$ with a specified $U_\infty$ value.  For the $U'_F$
surface boundary condition from (\ref{bc}), this implies that
\begin{equation}
  U'_{F}(0) \,=\, -\, (U_\infty - U_*) \,.
\label{UFbc}
\end{equation}
Then $F'(Z)$ is chosen such that this condition is satisfied.  Thus, we can
specify a simple representative shape,
\begin{equation}
  F'(Z) \,=\, F_0 \, e^{-\, \ell Z } \,,
  \label{Fshape}
\end{equation}
whose amplitude $F_0$ is chosen to satisfy (\ref{UFbc}).  The analytic
solution for $\ell \ne q$ is
\begin{align}
  \psi'_F(Z) &=\, \psi_0 \,\left(\, e^{-\ell Z} \,-\, \cos[qZ] \,e^{-qZ} \,+\,
  \left(\frac{\ell}{q} - 1\right)\, \sin[qZ] e^{-qZ} \,\right)
  \nonumber \\
  b'_F(Z) &=\, \,-\, \frac{N^2\sin[\theta]\psi_0}{\ell \kappa} \,\left(\,
  e^{-\ell Z} \,-\, \frac{\ell}{q}\, \cos[qZ] \,e^{-qZ}  \,+\,
  \frac{\ell^2}{2 q^2}\, (\sin[qZ]+\cos[qZ])\, e^{-qZ} \,\right) 
  \nonumber \\
  U'_F(Z) &=\, \left( \frac{f\cos[\theta]\psi_0}{\ell\nu} \,-\,
  \frac{F_0}{\ell^2\nu} \right)\,e^{-\ell Z} \,-\,
  \frac{f\cos[\theta]\psi_0}{q\nu}\,\cos[qZ]\, e^{-qZ}
  \nonumber \\
  &\qquad \,+\,
  \frac{f\ell\cos[\theta]\psi_0}{2q^2\nu}\,(\sin[qZ]+\cos[qZ])\, e^{-qZ}
  \nonumber \\
  V'_F(Z) &= \psi_0 \,\left(\, -\,\ell\,e^{-\ell Z} \,+\, 2q\sin[qZ] \,e^{-qZ} \,+\,
  \ell\,(\cos[qZ] \,-\, \sin[qz])\, e^{-qZ} \,\right) \,,
\end{align}
where
\begin{align}
  \psi_0 &=\, \frac{f\cos[\theta]}{\ell\nu^2(\ell^4 + 4q^4)} \ F_0
  \qquad \mathrm{and}
  \nonumber \\
  F_0 &=\, (U_\infty \,-\, U_*)\,\Big/ \left(\, \frac{1}{\ell^2\nu} \,-\,
    \frac{f^2\cos^2[\theta]}{\ell^2\nu^3}\,
    \left(\frac{1 - \ell/q + \ell^2/(2q^2)}{\ell^4 + 4q^4}\right)
  \,\right) \,.
\end{align}
With the parameter values listed in the caption to
Fig.~\ref{fig:pbUV}, $F_0 = -\, 5.8 \times 10^{-6}$ m s$^{-2}$ and
$\psi_0 = -\, 0.80$ m$^{2}$ s$^{-1}$ for the case referred to as (2)
in Sec.~\ref{sec:1D}.

\bibliographystyle{ametsocV6}
\bibliography{references}

\end{document}